\documentclass[12pt]{article}
\usepackage{mathrsfs}
\usepackage[T1]{fontenc}
\usepackage{mathpazo}
\usepackage{setspace}
\usepackage{amsfonts}
\usepackage{amssymb}
\usepackage{epsfig}
\usepackage{latexsym}
\usepackage{color}
\usepackage{graphicx}
\usepackage{nicefrac}
\usepackage[latin1]{inputenc}
\usepackage{slashed}
\usepackage{multirow}

\usepackage{hyperref}

\def\hybrid{\topmargin -30pt    \oddsidemargin 0pt %%%%%%%%%%%%%% Archive-30pt
        \headheight 0pt \headsep 0pt
        \textwidth 6.25in       % A4 paper
        \textheight 9.5in       % A4 paper
        \marginparwidth .875in
        \parskip 5pt plus 1pt   \jot = 1.5ex}

\hybrid

\def\baselinestretch{1.2}

\catcode`\@=11

\def\marginnote#1{}
\newcount\hour
\newcount\minute
\newtoks\amorpm
\hour=\time\divide\hour by60
\minute=\time{\multiply\hour by60 \global\advance\minute by-\hour}
\edef\standardtime{{\ifnum\hour<12 \global\amorpm={am}%
        \else\global\amorpm={pm}\advance\hour by-12 \fi
        \ifnum\hour=0 \hour=12 \fi
        \number\hour:\ifnum\minute<10 0\fi\number\minute\the\amorpm}}
\edef\militarytime{\number\hour:\ifnum\minute<10 0\fi\number\minute}
\def\draftlabel#1{{\@bsphack\if@filesw {\let\thepage\relax
   \xdef\@gtempa{\write\@auxout{\string
      \newlabel{#1}{{\@currentlabel}{\thepage}}}}}\@gtempa
   \if@nobreak \ifvmode\nobreak\fi\fi\fi\@esphack}
        \gdef\@eqnlabel{#1}}
\def\@eqnlabel{}
\def\@vacuum{}
\def\draftmarginnote#1{\marginpar{\raggedright\scriptsize\tt#1}}

\def\draft{\oddsidemargin -.5truein
        \def\@oddfoot{\sl preliminary draft \hfil
        \rm\thepage\hfil\sl\today\quad\militarytime}
        \let\@evenfoot\@oddfoot \overfullrule 3pt
        \let\label=\draftlabel
        \let\marginnote=\draftmarginnote
   \def\@eqnnum{(\theequation)\rlap{\kern\marginparsep\tt\@eqnlabel}%
\global\let\@eqnlabel\@vacuum}  }

\def\draft2{
        \def\@oddfoot{\sl preliminary draft \hfil
        \rm\thepage\hfil\sl\today\quad\militarytime}
        \let\@evenfoot\@oddfoot \overfullrule 3pt
        \let\label=\draftlabel
        \let\marginnote=\draftmarginnote
   \def\@eqnnum{(\theequation)\rlap{\kern\marginparsep\tt\@eqnlabel}%
\global\let\@eqnlabel\@vacuum}  }

\def\preprint{\twocolumn\sloppy\flushbottom\parindent 2em
        \leftmargini 2em\leftmarginv .5em\leftmarginvi .5em
        \oddsidemargin -.5in    \evensidemargin -.5in
        \columnsep .4in \footheight 0pt
        \textwidth 10.in        \topmargin  -.4in
        \headheight 12pt \topskip .4in
        \textheight 6.9in \footskip 0pt
        \def\@oddhead{\thepage\hfil\addtocounter{page}{1}\thepage}
        \let\@evenhead\@oddhead \def\@oddfoot{} \def\@evenfoot{} }

\def\numberbysection{\@addtoreset{equation}{section}
        \def\theequation{\thesection.\arabic{equation}}}

\def\underline#1{\relax\ifmmode\@@underline#1\else
        $\@@underline{\hbox{#1}}$\relax\fi}

\def\titlepage{\@restonecolfalse\if@twocolumn\@restonecoltrue\onecolumn
     \else \newpage \fi \thispagestyle{empty}\c@page\z@
        \def\thefootnote{\fnsymbol{footnote}} }

\def\endtitlepage{\if@restonecol\twocolumn \else \newpage \fi
        \def\thefootnote{\arabic{footnote}}
        \setcounter{footnote}{0}}  %\c@footnote\z@ }

\catcode`@=12
\relax

\def\figcap{\section*{Figure Captions\markboth
        {FIGURECAPTIONS}{FIGURECAPTIONS}}\list
        {Figure \arabic{enumi}:\hfill}{\settowidth\labelwidth{Figure
999:}
        \leftmargin\labelwidth
        \advance\leftmargin\labelsep\usecounter{enumi}}}
 \relax
\def\tablecap{\section*{Table Captions\markboth
        {TABLECAPTIONS}{TABLECAPTIONS}}\list
        {Table \arabic{enumi}:\hfill}{\settowidth\labelwidth{Table
999:}
        \leftmargin\labelwidth
        \advance\leftmargin\labelsep\usecounter{enumi}}}
 \relax
\def\reflist{\section*{References\markboth
        {REFLIST}{REFLIST}}\list
        {[\arabic{enumi}]\hfill}{\settowidth\labelwidth{[999]}
        \leftmargin\labelwidth
        \advance\leftmargin\labelsep\usecounter{enumi}}}
 \relax
\makeatletter
\newcounter{pubctr}
\def\publist{\@ifnextchar[{\@publist}{\@@publist}}
\def\@publist[#1]{\list
        {[\arabic{pubctr}]\hfill}{\settowidth\labelwidth{[999]}
        \leftmargin\labelwidth
        \advance\leftmargin\labelsep
        \@nmbrlisttrue\def\@listctr{pubctr}
        \setcounter{pubctr}{#1}\addtocounter{pubctr}{-1}}}
\def\@@publist{\list
        {[\arabic{pubctr}]\hfill}{\settowidth\labelwidth{[999]}
        \leftmargin\labelwidth
        \advance\leftmargin\labelsep
        \@nmbrlisttrue\def\@listctr{pubctr}}}
 \relax
\makeatother

\newcommand{\nn}{\nonumber \\}

\def\be{\begin{equation}}
\def\ee{\end{equation}}
\def\ba{\begin{eqnarray}}
\def\ea{\end{eqnarray}}

\def\del{\partial}

\def\r{\rho}
\def\a{\alpha}

\def\b{\beta}

\def\G{\Gamma}
\def\d{\delta}

\def\e{\epsilon}

\def\th{\theta}

\def\m{\mu}
\def\n{\nu}
\def\om{\omega}
\def\Om{\Omega}

\def\L{\Lambda}
\def\s{\sigma}

\def\cL{{\cal L}}

\def\no{\noindent}

\def\qq{\qquad}

\def\IR{\relax{\rm I\kern-.18em R}}

\def\vol{{\rm Vol}}

\def\inv{^{\raise.0ex\hbox{${\scriptscriptstyle -}$}\kern-.05em 1}}

\def \ha {{\frac{1}{2}}}

\def \ov {\over}

\begin{document}
%\draft2

%\renewcommand{\theequation}{\arabic{equation}}
%\renewcommand{\theequation}{\thesection.\arabic{equation}}

\renewcommand{\theequation}{\thesection.\arabic{equation}}
\csname @addtoreset\endcsname{equation}{section}

\begin{titlepage}
\begin{center}

\hfill FPAUO-12/08  \\

\phantom{xx}
\vskip 0.4in

%{\large \bf  Consistency of non-Abelian T-duality from seven-dimensions}

%{\large \bf  A lower dimensional view point of non-Abelian T-duality in string theory }

%{\large \bf  Lower dimensional view of non-Abelian T-duality\\ in type-II supergravity}

{\large \bf  Non-Abelian T-duality and consistent truncations \\ in type-II supergravity}

\vskip 0.4in

{\bf Georgios Itsios}${}^{1,3 a}$,\phantom{x} {\bf Yolanda  Lozano}${}^{2b}$,
\\

\vskip .15 cm

{\bf Eoin  \'O Colg\'ain}${}^{2c}$\phantom{x}  and \phantom{x} {\bf Konstadinos Sfetsos}${}^{1d}$ \vskip 0.1in

\vskip .2in

${}^1$Department of Engineering Sciences, University of Patras,\\
26110 Patras, Greece\\

\vskip .2in

${}^2$Department of Physics,  University of Oviedo,\\
Avda.~Calvo Sotelo 18, 33007 Oviedo, Spain\\

\vskip .2in

${}^3$ Centro de Fisica do Porto \& Departamento de Fisica e Astronomia,\\
Faculdade de Ciencias da Universidade do Porto, Rua do Campo Alegre 687, 4169-007 Porto, Portugal

\vskip .2in

\end{center}

\vskip .4in

\centerline{\bf Abstract}

\no
For a general class of $SO(4)$ symmetric backgrounds in type-II supergravity, we show that the action of
non-Abelian T-duality can
be described via consistent truncation to seven dimensional theories with seemingly massive modes.
As such, any solution to these theories uplifts to both massive
type IIA and IIB supergravities presenting an invertible map between the two.
For supersymmetric backgrounds, we show that for spinors transforming under $SO(4)$ non-Abelian T-duality breaks
the original supersymmetry by half. We use these mappings to generate the
non-Abelian T-duals of the maximally supersymmetric pp-wave,
the Lin, Lunin, Maldacena geometries and spacetimes with Lifshitz symmetry.

\vfill
\no
 {%\footnotesize
$^a$gitsios@upatras.gr,
 $^b$ylozano@uniovi.es,
$^c$ocolgain@gmail.com,
$^d$sfetsos@upatras.gr.}

\end{titlepage}
\vfill
\eject

%\def\baselinestretch{1.2}
%\baselineskip 10 pt
%\noindent

\tableofcontents

\def\baselinestretch{1.2}
\baselineskip 20 pt

\newcommand{\eqn}[1]{(\ref{#1})}

\section{Introduction}

Recently our understanding of non-Abelian T-duality \cite{Fridling:1983ha,Fradkin:1984ai,delaossa:1992vc}
has been considerably advanced by showing how to
implement the duality transformation on solutions of type-IIA (massive) and type-IIB supergravities with
non-trivial RR fluxes and a non-Abelian group of isometries \cite{Sfetsos:2010uq}.
Originally the formulation was
implemented on supergravity backgrounds in which isometries were realized
with group spaces \cite{Sfetsos:2010uq}. Nevertheless, it was soon extended
to cover backgrounds in which the isometries were realized with coset spaces  \cite{Lozano:2011kb} as in the vast majority
of interesting solutions appearing in supergravity and in string theory.

\no
In the present paper we put on firmer ground the previous work by focusing not
on particular supergravity solutions but on the corresponding massive IIA and type-IIB
supergravity theories themselves. As a playground we choose a
general class of $SO(4)$ symmetric backgrounds
and examine non-Abelian T-duality with respect to an $SU(2)$
subgroup.\footnote{Neglecting a $B$-field with field strength $H$ along $S^3$, this is the most general ansatz.}
We unearth consistent reduction ansatze to the same underlying seven dimensional theories meaning that \textit{any}
solution to these lower-dimensional actions uplifts simultaneously to a solution of both
type-IIA and type-IIB supergravities.
Some of our motivation lends itself to the important work of Bergshoeff, Hull \& Ort\'in \cite{Bergshoeff:1995as} where
the Abelian T-duality rules are derived via simultaneous circle reductions and a subsequent matching of the fields using the fact that the $\mathcal{N}=2$
nine-dimensional supergravity is unique.

\no
Despite the casual analogy, in venturing from circle reductions to geometries connected via non-Abelian T-duality,
one has an important obstacle to clear; non-Abelian T-duality breaks isometries, and in the particular setting of this paper,
 one has to compare an $S^3$ reduction of the original geometry with an $\mathbb{R} \times S^2$ reduction in the T-dual.
Indeed, as we shall see, delicate cancellations have to happen where the Einstein equations along $S^3$ get mapped to those
along  $\mathbb{R} \times S^2$ and the $B$-field equation in the T-dual conspires to mimic this result.
Another potential surprise may be that the non-Abelian transformation of the original RR flux ansatz leads one  to a T-dual
reduction ansatz that reproduces the \textit{same} lower-dimensional theory.  A priori, it is not obvious that some reshuffling
of the field content may not be required. However, this theory, which arises as an $S^3$ reduction from type-IIB with $SO(4)$ singlets retained,
does not agree with the Lagrangian expected from a (warped) $S^3$ reduction from type-IIB \cite{Samtleben:2005bp},
since although some fields, such as the dilaton, fit into the massless supergravity multiplet and conform to expectations,
the warp factor and the axion, the other $SO(4)$-singlet scalars of the reduction, appear to be massive.

\no
Recall that the identification of consistent dimensional reductions can be a conceptually and technically demanding task.
Nevertheless, our most celebrated examples of consistent truncations involve maximally symmetric
sphere reductions \cite{s51}-\cite{Nicolai:2011cy} where there are no obvious guidelines to constructing a consistent reduction ansatz.
In contrast, consistent truncations are relatively easy to work out when one
has generic $SU(3)$-structure \cite{KashaniPoor:2007tr}-\cite{Colgain:2010rg}
or $SU(2)$-structure \cite{Cassani:2010uw}-\cite{Cassani:2011fu} manifolds
allowing the possibility to expand in the invariant forms.\footnote{For the reduction of fermions see \cite{Bah:2010yt}.} In addition, new reduction ansatze may sometimes be
deduced from known ansatze, as in \cite{Cvetic:2000ah}, where an $S^3$ reduction in type IIA is derived as
a limit of the $S^4$ reduction of \cite{s41} involving the $S^4$ pinching off to $ \mathbb{R} \times S^3$,
or alternatively via Abelian T-duality \cite{Colgain:2011ng}. In this paper, the underlying principle
illuminating the reduction on the T-dual spacetime with factor $\mathbb{R} \times S^2$ is non-Abelian T-duality.

\no
Non-Abelian T-duality breaks isometries, and as such, it is expected that supersymmetry
is also broken.\footnote{This is correct at the supergravity level. In string theory supersymmetry can be realized
in a non-local way (see \cite{Sfetsos:1996pm} and references therein).}
From our extensive knowledge of Abelian T-duality in type-II supergravity, we know that the Killing spinor
equations of the original solution can be mapped to the Killing spinor equations of the T-dual via a rotation acting
exclusively on one of the chiral Killing spinors, and that in the process, the chirality of the spinor flips \cite{Hassan:1999bv}.
Here we generalise this by showing that there is an analogous rotation when we perform an $SU(2)$ transformation and that the Killing spinor
equations of the T-dual are those of the original up to the imposition of a single further supersymmetry variation. For spinors transforming under $SO(4)$ 
we show that this additional condition is consistent with half of the supersymmetries, those corresponding to the $SU(2)$ isometry
with respect to which we perform the non-Abelian T-duality, breaking.

\no
The organization of the paper is as follows: In section 2 we present the general class of type-II $SO(4)$ symmetric backgrounds
that we T-dualize with respect to an $SU(2)$ subgroup. Focusing on type-IIB we reduce both the original IIB background (on $S^3$)
and its massive type-IIA dual (on $ \mathbb{R} \times S^2$) and obtain a consistent truncation in seven dimensions, both at the level of
the equations of motion and at the level of the actions. We compare the resulting seven dimensional action to previous constructions of maximal
supergravities in the literature. In section 3 we analyze the supersymmetry of the non-Abelian dual. We show that the mapping of
the Killing spinor equations requires an additional condition, breaking the original supersymmetry by a half for spinors transforming under $SO(4)$. In section 4 we focus
on massive type-IIA $SO(4)$ symmetric backgrounds and show that the reduction on $S^3$ produces as well a consistent truncation which
is however different from the one obtained in the reduction of type-IIB on $S^3$. In section 5 we present three examples in type-IIB in which our non-Abelian T-duality
transformation can be used to generate new solutions of massive IIA. These are the maximally supersymmetric
pp-wave background \cite{Blau:2001ne}, the general class of 1/2 supersymmetric type-IIB solutions constructed in \cite{LLM}
and the Lifshitz solutions of \cite{Balasubramanian:2010uk}. We also exhibit the reverse transformation from type-IIA to type-IIB using the pp-wave.
The non-Abelian dual of mass deformed ABJM provides an additional example in type-IIB very similar in nature to the dual of the 1/2 supersymmetric solutions in \cite{LLM}.
Section 6 contains our conclusions and further directions. Appendix A summarizes some aspects of type-II supergravities relevant
to our work. Appendix B contains the details of the reductions of type-IIB on $S^3$, type-IIA on $\mathbb{R} \times S^2$
and type-IIA on $S^3$. Finally, Appendix C contains the details of the derivation of the dual Ramond fields, which is used as an ansatz in section 2.

\section{Consistent KK Reduction}
\label{sec:IIBduals}

In this section, for concreteness, we confine ourselves to spacetimes with warped $S^3$
factors permitting a non-Abelian $SU(2)$ transformation.
From a technical point of view since we will be dealing with a group manifold the isometry
acts with no fixed points, a fact that, as in \cite{Sfetsos:2010uq}, facilitates the
computations.
However, we expect that the picture we paint here will generalise to the
examples presented in \cite{Lozano:2011kb} where one encounters
larger isometry groups based on coset spaces, such as spheres, and the corresponding
T-duals have less supersymmetry and fewer isometries.

\no
As stated, we are interested in spacetimes of the form
\be
\label{origbac}
ds^2 = ds^2(M_7) + e^{2 A} ds^2(S^3)\ ,
\ee
where $M_7$ is a seven-dimensional Minkowskian spacetime and the warp factor $A$
depends only on the coordinates on $M_7$. The $S^3$ metric is normalized so that $R_{ij}= \ha g_{ij}$.
The above metric has an $SO(4)$ group of isometries, together with the
isometries of the $M_7$ manifold in which we will not be interested in our general discussion.
The NS-sector fields are comprised also by a two-form
$B$ with field strength  $H=dB$ with no-components along $S^3$, as well as
a dilaton $\Phi$ which may depend on the coordinates of $M_7$. Consequently, these fields
are also invariant under the $SO(4)$ isometry group.
Incorporating a $B$-field along $S^3$ will lead to a generalisation
which falls outside our scope in this paper.
Irrespective of the chirality of the theory,
one can write the $SO(4)$ as $SU(2)_{L} \times SU(2)_{R}$
and perform an $SU(2)$ transformation with respect to one of these factors as explained
in detail in \cite{Sfetsos:2010uq}.
The end result is a spacetime with fields in the NS-sector given by
\ba
&&  d\hat s^2 =ds^2(M_7) + e^{-2 A} dr^2 + {r^2 e^{2 A}\ov r^2 + e^{4 A}} ds^2(S^2)\ ,
\nonumber\\
&& \hat B =  B+  \tilde B \ , \qq \tilde B =  \frac{r^3}{r^2 + e^{4 A}} {\rm Vol}(S^2)\ ,
\label{NSsec}
\\
&& e^{-2 \hat \Phi} = e^{-2 \Phi} e^{2 A} (r^2 + e^{4 A})\ ,
\nonumber
\ea
where we have used hat notation to differentiate the T-dual fields from the original ones.
Observe also that the $SU(2)$ isometry left untouched by the transformation is captured
in the symmetries of the resulting $S^2$.
The field strength, $\hat H = d\hat B$ may be written as
\be
\hat H = H +  \left[\frac{r^2(r^2 + 3 e^{4 A})}{(r^2+e^{4 A})^2} dr
- 4 \frac{r^3  e^{4 A}}{(r^2 + e^{4 A})^2} d A \right] \wedge {\rm Vol}(S^2) \ .
\ee
To construct solutions of type-II supergravity, we need to complement our original
spacetime ansatz (\ref{origbac}), with knowledge of the RR fields.
Building on the tradition started in \cite{Sfetsos:2010uq},
and the two known examples which will serve as valuable consistency checks, we begin by considering first
type-IIB supergravity.
Through the existence of the known examples, we know that our $SU(2)$ transformation takes solutions of
the equations of motion in type-IIB
to solutions in massive IIA, strongly suggesting that by examining the equations of motion  one
can unearth some deeper structure.
Indeed, as we shall see shortly, this is the case and the underlying structure that emerges is a
unifying gravity description in seven dimensions
via parallel consistent truncations on the original spacetime \eqn{origbac} and on the T-dual
spacetime (\ref{NSsec}).
As we shall see, in this formulation non-Abelian
T-duality is invertible unlike the case of the standard $\sigma$-model approach.\footnote{
The transformation is invertible also in the context of Poisson--Lie T-duality in which a non-standard $\s$-model action
lacking manifest Lorentz invariance is used \cite{Klimcik:1995dy}.}

\subsection{KK reduction on $S^3$}

Our first task is to identify the reduced seven dimensional theory.
To do this we incorporate into our type-IIB ansatz the following RR fluxes
that respect the symmetry of the round $S^3$ appearing in the metric
\ba
&& F_5 = G_2 \wedge {\rm Vol}(S^3) - e^{-3 A} \star_7 G_2\ ,
\nonumber\\
&& F_3 = G_3 - m {\rm Vol}(S^3)\ ,
\label{IIBfluxansatz}
\\
&& F_1 = G_1\ .
\nonumber
\ea
Note that the self-duality of the five-form has already been imposed according to our conventions, see
\eqn{defhod} below.
We take the forms $G_i$, $i=1,2,3$ to live on $M_7$.  Thanks to the type-IIB Bianchi identities (see Appendix A.1)
the parameter $m$ is a constant and as it turns out, it will be mapped to the mass parameter of the
massive IIA supergravity. Aspects of type-II supergravities relevant to this work are reviewed in Appendix A.

\no
The consistency of this reduction should come as no surprise and we have indeed checked
that one can get the same result
by performing the reduction at the level of the action. We enclose details of the
type-IIB reduction in the Appendix B.1.
In particular, the reduction implies that
the forms we used in our ansatz can be expressed in terms of some potentials as
\be
G_1 = dC_0 \ ,\qq G_2=   dC_1 - m B \ ,\qq G_3 = dC_2 -C_0 H \ .
\label{fielstreIIB}
\ee
Hence the field content arising from the RR sector is a scalar, $C_0$, a one-form,
$C_1$ and a two-form, $C_2$.
These supplement the metric, the two-form $B$-field and the dilaton $\Phi$ from the NS sector.
All these arise from the effective action which is given in the Einstein frame by \eqn{EEEUN}.
After performing the transformation
\be
\Phi = 5 \tilde{\Phi} + \frac{3}{2} A\ ,
\label{reddf}
\ee
which allows us to diagonalize the action and compare dilaton factors directly with the action (6.20) in \cite{Samtleben:2005bp},
we arrive, dropping the tildes, at the final action in Einstein frame
from whence all equations of motion may be derived. The corresponding Lagrangian density is
\ba
\label{D7action}
&& \mathcal{L}_{\rm Einstein} =  {R}
- 3 (\partial A)^2 - 20 (\partial \Phi)^2 - \frac{1}{2} e^{10 \Phi + 3A}(\partial C_0)^2
-\frac{1}{12} e^{ - 8 \Phi} H^2
\nn
&& \phantom{xxxxxxxx} - \frac{1}{2} \left( m^2 e^{ 14 \Phi-3 A}
- 3 e^{ 4 \Phi -2 A }  +  \frac{1}{2} e^{ 6 \Phi-3 A} G_2^2
+ \frac{1}{6} e^{2 \Phi + 3 A} G_3^2\right)
\\
&& \phantom{xxxxxxxx} +  G_2 \wedge C_2 \wedge H\ ,
\nonumber
\ea
where we have made use of \eqn{fielstreIIB} and have replaced $G_1$ by $dC_0$.

\no
Now that we have the seven-dimensional action in Einstein frame we
can attempt to make contact with the supergravity literature. The warped $S^3$
reduction ansatz from type-IIB is still unknown, but various reductions from
type-I supergravity have been discussed, notably the reduction ansatz of  \cite{Chamseddine:1999uy}, 
which through the ten-dimensional equations of motion, reproduces the equations of motion of \cite{Townsend:1983kk},
 and the ansatz of \cite{Cvetic:2000dm} which leads to \cite{Salam:1983fa}.
More generally, in seven dimensions one can construct maximal
supergravities \cite{Samtleben:2005bp} (see section 6.3) generalising \cite{Cvetic:2000dm, Salam:1983fa}.
 Attempts to match our action to the general action of \cite{Samtleben:2005bp}
reveal that only the dilaton factors agree perfectly while neither our warp factor,
 $A$,  nor the axion, $C_0$, fit into this work.

\no
The expectation then is that $A$ and $C_0$ correspond to scalars
in massive multiplets. As our potential has no stationary points, determining
the mass of these terms relative to the dilaton becomes tricky. In spite of these difficulties,
 the supergravity spectrum for warped $S^3$ solutions corresponding to the near-horizon geometry
of D5-branes may be found in table IV of \cite{Morales:2004xc}. One observes that in the
full spectrum there are three $SO(4)$ singlet scalars, i.e. representation [00](000),
 which show up in three different multiplets: $n=2$, $n=3$ and $n=4$.
Now only the $n=2$ multiplet corresponding to the dilaton is in the massless
supergravity multiplet, while the other two are massive.\footnote{We are grateful to H. Samtleben
for correspondence on this point.} This suggests that more generally $A$ and $C_0$
are massive modes and that their omission from the maximal
supergravity action should not come as a surprise.

\no
We also note that the overall coupling constant of the seven-dimensional theory in terms of the coupling
constant of type-IIB supergravity is
\be
{1\ov 2 \kappa_7^2}= {\vol(S^3)\ov 2 \kappa^2}\ .
\ee

\no
In addition it is a lengthy but otherwise straightforward procedure to demonstrate that
by dimensionally reducing the type-IIB supergravity action \eqn{actIIB}, then passing to the
Einstein frame and finally by redefining as in \eqn{reddf} we obtain
precisely the action \eqn{D7action}. We have decided not to include the details of this calculation
as it presented no technical or conceptual challenges.

\subsection{Non-Abelian T-duality and KK reduction}
We have consistently reduced the general $SO(4)$ invariant ansatz of type-IIB down to seven dimensions
at the level of the equations of motion
and noted that this is also possible at the action level.
The question we would like to address now is whether such a reduction will be possible for the
non-Abelian T-dual background with respect to an $SU(2)\subset SO(4)$.
Indeed, our knowledge of non-Abelian T-duality in this setting is confined
to two solitary examples constructed in \cite{Sfetsos:2010uq} involving the near horizon geometry of
the D1-D5- and the D3-brane systems
corresponding to the $AdS_3\times S^3\times T^4$ and the $AdS_5\times S^5$
geometries.\footnote{More examples
were constructed in \cite{Lozano:2011kb} involving coset and not group spaces.}
So it may even be too much to expect that there is an overarching action in seven dimensions
describing the full reduction and not just separate actions
corresponding to a truncation to $m, G_3$ (D1-D5 near-horizon) and $A, G_2$ (D3 near-horizon) or
$G_1$ separately.
The surprise, as we shall see shortly,
is that one obtains exactly the same theory in seven dimensions.

\no
The form of the RR flux fields can certainly be constrained by the symmetries of the non-Abelian T-dual.
It is apparent from the expressions for the NS sector \eqn{NSsec} that in the non-Abelian T-dual
backgrounds the $SO(4)$ isometry group is broken down to $SO(3)\sim SU(2)$, i.e. the symmetry group of $S^2$.
Hence, we have two natural forms to build an RR flux reduction ansatz from,  namely $dr$ and $\vol(S^2)$.
In type-IIA supergravity, decomposing the forms one has the natural ansatz
\ba
\label{schem}
\hat F_{2} &=& M_0 \vol(S^2) + M_1 \wedge dr + M_2 \ ,
\nn
\hat F_{4} &=&  N_1 \wedge dr \wedge \vol(S^2) + N_2 \wedge \vol(S^2) + N_3 \wedge dr + N_4\ ,
\ea
where $M_i$, $N_i$ denote forms of degree $n$ living on $M_7$.
The difficulty arises from the fact that all forms on $M_7$ in the above ansatz can still
depend on the radial direction $r$.
One approach then is to employ trial and error and match the equations of
motion of massive IIA supergravity to those of type-IIB,
so that the $M_i$ and $N_i$ align with our $G_i$ from the type-IIB reduction discussed previously.
While this approach may reap a reward if one just focuses on reproducing the equations of motion from type-IIB reduced on $S^3$,
i.e. if one puts the answer in by hand, it is difficult to find a general reduction ansatz with cohomogeneity-one manifolds (for example, see \cite{Colgain:2011ng}).
%This procedure will result into a system of partial differential equations
%and clearly is not either practical nor instructive.
Alternatively, from earlier work \cite{Sfetsos:2010uq} it is known how the fluxes transform,
so we can simply generate the appropriate ansatz using the type-IIB flux ansatz as a seed.
Obviously this is the preferred approach which we follow.

\no
Firstly, one constructs the type-IIB flux bispinor from the ansatz (\ref{IIBfluxansatz})
\be
P ={e^\Phi\ov 2}\sum_{n=0}^4 {\slashed{F}_{2 n+1} \ov (2n+1)!}\ ,
\label{bisp1}
\ee
where we have employed the usual notation $\slashed{F}_p \equiv  F_{i_1\dots i_p} \G^{i_1 \dots i_p}$,
and reads off the T-dual bispinor from the transformation
\be
\hat{P} =  P\Omega^{-1}\ ,
\label{ppom}
\ee
where $\Omega$ is the Lorentz transformation matrix acting on the spinors. It reads \cite{Sfetsos:2010uq}
\be
\Om = \G_{11} {e^{2A} \G_{789} + {\bf x} \cdot {\bf \G}\ov \sqrt{r^2 + e^{4 A}}}\qq \Longrightarrow \quad
\Om^{-1} = \G_{11} {e^{2A} \G_{789} - {\bf x} \cdot {\bf \G}\ov \sqrt{r^2 + e^{4 A}}}\ .
\label{omega}
\ee
Note that we are using $e^{i}$, $i=7,8,9$ to denote the tangent space along the transformed T-dual space.
A natural choice of frame may also be found in \cite{Sfetsos:2010uq}
\be
\hat e^i = {1\ov  \sqrt{r^2 + e^{4 A}}} \left(e^A dx^i + x^i  e^{-A} b(r) dr\right)\ , \qq
b(r) =  {\sqrt{r^2 + e^{4 A}} -e^{2 A}\ov r}\ .
\label{fraamwe}
\ee
Our conventions on Hodge duality are such that on a
$p$-form in a $D$-dimensional spacetime
\be
(\star F_p)_{\m_{p+1}\cdots \m_D}  = {1\ov p!} \sqrt{|g|}\ \e_{\m_1\cdots \m_D} F_p^{\m_1\cdots \m_p}\ ,
\label{defhod}
\ee
where $\e_{01\dots 9} =1$.  With this we have the useful identity $\star\star F_p = s (-1)^{p(D-p)}
F_p$, where $s$ is the signature of spacetime which we take to be mostly plus.
In our case, the indices, $\mu=0,1,\dots , 6$ refer to the seven-dimensional spacetime $M_7$
of Minkowski signature,
whereas $7,8,9$ either to the $S^3$ directions or, for the non-Abelian T-dual, to the frame defined in
\eqn{fraamwe}.

\no
The details of the construction are presented in Appendix C.1.
The final form of the fluxes may then be read off from the T-dual bispinor
\be
\hat P ={e^{\hat \Phi}\ov 2}\sum_{n=0}^5 {\hat {\slashed{F}}_{2 n} \ov (2n)!}\ .
\label{bisp2}
\ee
This procedure gives the massive IIA fluxes, that we read from equation \eqn{flllhatIIB}
\ba
&& \hat F_0  =  m\ ,
\nonumber\\
 && \hat F_2 = \frac{m r^3}{r^2 + e^{4 A}} \vol(S^2) + r dr \wedge G_1 - G_2\ ,
\label{fliIIAmas}
\\
&&\hat F_4 = \frac{r^2 e^{4 A} }{r^2+e^{4A}} G_1\wedge dr \wedge \vol(S^2)
- \frac{r^3}{r^2 + e^{4 A}} G_2 \wedge \vol (S^2) +r  dr\wedge  G_3  + e^{3 A} \star_7 G_3 \ .
\nonumber
\ea
A quick inspection shows that they are of the form \eqn{schem}.
One can now use these fluxes in tandem with the T-dual spacetime (\ref{NSsec})
and plug them both into the massive IIA equations of motion.
We spare the reader the details but just summarize the necessary steps.
From the Bianchi identities and the flux equations of motion one recovers (\ref{biaiIIB}) and
the last two eqs. of (\ref{fluxIIA}).
From the $B$-field equation of motion one gets the first of (\ref{fluxIIA}) and
\ba
\label{Beom2}
&& \frac{1}{2} e^{A-2\Phi} \vol(M_7) - d \left(e^{3 A - 2 \Phi } \star_7 d A\right) = \frac{1}{4}
\biggl( m^2 e^{-3 A} - e^{3A} G_1 \wedge \star_7 G_1
\nn
&& \qq \qq\qq \phantom{xxxxxx}
+  e^{-3 A} G_2 \wedge \star_7 G_2- e^{3A} G_3 \wedge \star_7 G_3  \biggr)\ .
\ea
Observe that (\ref{Beom2}) is just (\ref{Eincons1}) written in a form notation.
The Einstein equations in the $r$-direction and
the directions along $S^2$ deliver (\ref{Eincons1}).
In all cases the dependence on the $r$-coordinate drops out.
Finally, one recovers the seven-dimensional Einstein equation (\ref{D7Ein}).

\no
Now that we have discussed two consistent reduction ansatze from type-II leading to the
same gravitational action in seven dimensions, we pause to recap on what we have shown. Notably, any solution to the action (\ref{D7action})
may be uplifted simultaneously to both type-IIB and
massive IIA.\footnote{Another prominent example of non-unique uplifts
of solutions to higher dimensions
includes solutions of the Romans theory in
five dimensions \cite{romans5d}. These can be uplifted to type-IIB supergravity
\cite{lpt} and to eleven-dimensional supergravity \cite{5dto11d,5dto11dgen}. However, unlike our case,
there is no obvious relation between these distinct uplifts.} Therefore, in this formulation we have circumvented
the problem of non-invertibility of non-Abelian T-duality in its treatment in the standard two-dimensional $\s$-model
approach, namely the fact that it is not possible to reconstruct the original background from its non-Abelian
dual due to its lack of isometries.
In addition, starting with a type-IIB geometry with an $S^3$, non-Abelian T-duality will generate a background to massive
type-IIA whenever the original $F_3$-flux extends on the $S^3$ directions (see eq. (\ref{IIBfluxansatz})).
Therefore non-Abelian duality can be used as a principle to generate new solutions to massive type-IIA supergravity.

%%%%%%%%%%%%%%%%%%%%%%
%\vskip 3 cm
%\begin{figure}[htp]
%\begin{center}
%\includegraphics [height=15 cm]{T-dual_Sch1.pdf}
%\end{center}
%\vskip -1 cm
%\caption{Schematic relation of the reduced seven-dimensional theories to type II supergravities.}
%\label{fig-ratiosphere}
%\end{figure}
%%%%%%%%%%%%%%%%%

%%%%%%%%%%%%%%%%%%%%
%\vskip 0 cm
%\begin{figure}%[!bh]
%\centering
%\includegraphics[scale= .55]{T-dual_Sch1.pdf}
%\vskip -2 cm
%\caption{Schematic relation of the reduced seven-dimensional theory to type II supergravities.}
%\label{fig1}
%\end{figure}
%%%%%%%%%%%%%%%%%%%%%%%%%

\subsection{Reduction at the level of the action}

It has already been pointed out that the seven-dimensional action \eqn{D7action},
besides giving rise to the same equations of
motion following either from the general $SO(4)$ invariant ansatz \eqn{origbac} and (\ref{IIBfluxansatz})
or from its non-Abelian dual (\ref{NSsec}) and \eqn{fliIIAmas}, also arises from the dimensional reduction
of the type-IIB action on $S^3$. Conceptually, it is not obvious that a dimensional reduction of massive IIA
on the T-dual background with topology $\mathbb{R} \times S^2$ will result into the same action \eqn{D7action},
or at least, if that is the case, it should happen in a non-trivial way.
Unlike the case of the usual dimensional reduction on compact manifolds where
the zero mode in a harmonic expansion
of the various fields is retained, the non-Abelian T-dual background seems to be non-compact
due to the fact that the radial coordinate $r$ seems to take values in the entire half real line.
In order to fully clarify the global topological properties of the dual background we should resort to
the $\sigma$-model derivation. However, how to extract topological information in the non-Abelian case is
not clear (see \cite{Alvarez:1993qi}). In any case in order to reproduce \eqn{D7action} 
it is key that the dual background has topology
$\mathbb{R} \times S^2$, as we show below.
Recall also that a treatment of non-Abelian T-duality of exact Conformal Field
Theory models in \cite{Polychronakos:2010hd} involving only
NS fields, led to the conclusion that non-Abelian T-duals effectively capture states of
some parent theory corresponding to
group theory representations with infinitely high highest weight.
Based on that work we expect that the dimensional reduction of non-Abelian
duals will capture this phenomenon.

\no
Keeping these in mind we start reducing the massive IIA action on the dual background.
All relevant terms are given in appendix B.2. Substituting them into \eqn{act2}
and after a partial integration, one obtains

\newpage
\ba
 S& = & \frac{\vol(S^{2})}{2\kappa^{2}}\int dr
 r^{2}\sqrt{-g}\left\{e^{3A-2\Phi}\left(R+6(\partial A)^2-12\partial A\cdot\partial\Phi+4(\partial\Phi)^{2}
\right.\right.
\nonumber\\
&& \phantom{xxxxxxx}
-\left.\frac{H^{2}}{12}+\frac{e^{-2A}}{2(r^2+e^{4A})^2}(3r^{4}+6r^{2}e^{4A}+8r^{2}e^{2A}+27e^{8A}) \right)
\nonumber\\
&& \phantom{xxxxxxx}
-\left.\frac{1}{2}\left(e^{-3A}m^{2}+e^{3A}G_{1}^{2}+\frac{e^{-3A}}{2}G_{2}^{2}+\frac{e^{3A}}{6}\frac{r^{2}
-e^{4A}}{r^{2}+e^{4A}}G_{3}^{2}\right)\right\}
\\
&&  \phantom{xxxxxxx}
-\frac{\vol(S^{2})}{2\kappa^{2}}\int dr\frac{r^4}{r^{2}+e^{4A}}C_{2}\wedge G_{2}\wedge H\wedge dr\ .
\nonumber
\ea
In the above action, there are divergent integrals with respect to $r$.
To deal with them we perform this integration from $0$ to $R_{0}$,
where $R_0$ is a cutoff which we take much larger than $e^{2A}$ while keeping as well the dominant term.
Then we can write the above action in the form \eqn{kgk22} but with overall coefficient
\be
\frac{\vol(S_{R_{0}}^{2})}{2\kappa^{2}} = {1\ov 2 \kappa_7^2}\ ,
\ee
where $\vol(S_{R_{0}}^{2})=\frac{4}{3}\pi R_{0}^{3}$ is the
volume of this large 2-sphere.
In order to keep the seven-dimensional Planck constant finite we have to
take the ten dimensional coupling constant $\kappa^2$ very large
as well so that the ratio is finite.
This is in resonance with the results of \cite{Polychronakos:2010hd}
where the overall
coupling constant of the theory  had to be taken large in order to
accommodate the consistent description of states corresponding to
infinitely large highest weight group theory representations.

\section{Supersymmetry}

In this section we examine the supersymmetry of the massive IIA T-dual
theory in relation to that of the original
type-IIB theory. In particular, we are interested in uncovering the mapping of
the corresponding Killing spinor equations. In addition,
we are after a general statement concerning the amount of supersymmetry
preserved under the non-Abelian T-duality transformation.
Based on the examples of  \cite{Sfetsos:2010uq}, where a key role was played by the Lie--Lorentz
or Kosmann derivative on spinors \cite{Kosmann,FigueroaO'Farrill:1999va,Ortin:2002qb}, 
we expect that for spinors transforming under $SO(4)$ supersymmetry is
reduced by half,
a statement that we actually prove below in this section.

\no
We will work with a consistent
set of conventions \cite{Hassan:1999bv} reproduced in Appendix A.3.

\no
It it instructive to consider first the non-Abelian T-dual of the flat spacetime metric
\be
ds^2 = -dt^2 + \cdots + dx_5^2 + dx^2 + {x^2\ov 4} ds^2(S^3) \ ,
\ee
without any RR fluxes, or other fields excited. Here the Killing spinor equation tells us
that the spinor is covariantly constant, namely $D_\m \e = 0$. As before the $S^3$ metric is normalized
such that $R_{ij}={1\ov 2} g_{ij}$. We conveniently write the $S^3$-metric as
\be
ds^2(S^3) = 4 (d\th^2 +\cos^2\th d\phi^2 + \sin^2\th d\psi^2)\ .
\label{ds32}
\ee
In this coordinate system the Killing spinor equation is solved by
\be
\e =\exp\left(\frac{\theta}{2} \G^{x \theta} \right)
\exp \left(\frac{\phi+\psi}{2} \G^{ x \phi} \right) \e_0\ ,
\label{flaski}
\ee
with $\e_0$ being a constant spinor and where we have used the natural orthonormal
frame suggested by the form of the metric. Indices in Gamma matrices belong to the tangent space.
The non-Abelian T-dual background is found from \eqn{NSsec} to be
\ba
ds^2 &=& -dt^2 + \cdots +  dx_5^2
+ dx^2 + \frac{4}{x^2} dr^2 + \frac{4 r^2 x^2}{16 r^2 + x^4} ds^2(S^2)\ ,
\nn
B &=& \frac{16 r^3}{16 r^2 + x^4} \vol(S^2)\ ,
\\
\Phi &=& - \frac{1}{2} \ln \left[ \frac{x^2}{4} \left( r^2 + \frac{x^4}{16} \right) \right]\ .
\nonumber
\ea
The dilatino variation in \eqn{killIIA} or equivalently \eqn{killIIB} is solved through the projector
\be
\label{proj1}
\G^{xr \theta \phi} \s^3 \e = -\e\ ,
\ee
where $\th$ and $\phi$ refer to the coordinates along $S^2$.
The gravitino variation of the same equations can also be readily solved by
\be
\label{ks1}
\e = \exp \left[ -\frac{1}{2} \tan^{-1} \left( \frac{1}{4} \frac{x^2}{r} \right)
\G^{\theta \phi} \s^3   \right]
 \exp \left(  \frac{\theta}{2} \G^{x \theta} \right) \exp \left( \frac{\phi}{2}
 \G^{\theta \phi} \right) \e_0 \ ,
\ee
where $\e_0$ denotes a constant spinor. We see that supersymmetry has been
broken by one half through the introduction of the projection condition in \eqn{proj1}.

\no
The above expression for the Killing spinor, after reintroducing the original warped factor
by replacing $\frac{x^2}{4}$ with $e^{2 A}$, suggests a substitution of the form
\be
\label{rotate}
\e =  e^{X} \tilde{\e} =  \exp \left( - \frac{1}{2} \tan^{-1} \left( \frac{e^{2A}}{r} \right)
\G^{\theta \phi} \s^3 \right) \tilde{\e}\ ,
\ee
into the Killing spinor equations of the original theory either in type-IIB or in massive IIA.
%\no
%{\bf Expression for $e^X$:} The conformal factor $e^X$ is identifiable with the $\Om$. This can be written
%\ba
%e^X = \Om & = &  {e^{2A}-(x_7\hat \G_{89} + {\rm cyclic\ in}\ 7,8,9)\ov \sqrt{r^2+e^{4A}}} \G_{789}\s_3
%\nonumber\\
%& = & \exp\left(\tan^{-1}\left(e^{2 A}\ov r\right) \G_{\th\phi}\right) \G_{r\th\phi}\s_3\ ,
%\ea
%where the indices $r,\th,\phi$ in the Gamma matrices are tangent space.
The above rotation is expected in the sense that a non-Abelian T-duality
results in a rotation on the spinor $\hat{\e} = \Omega \e$ \cite{Sfetsos:2010uq},
where $\Omega$ can be found in (\ref{omega}).
The rotation \eqn{rotate} should be accompanied by
a mechanism that changes the chirality of the theory. As we will see this will
involve the Gamma matrix $\G_r$
along the radial direction of the T-dual background.

\no
We next reintroduce the RR fluxes by first describing the
Killing spinor equations satisfied by the original type-IIB
background prior to the non-Abelian T-duality. Within the confines of our
ansatz (\ref{IIBfluxansatz}), the dilatino variation is\footnote{
All indices in the supersynmmetry variations below as well as in the Gamma matrices are,
unless otherwise stated, tangent space indices.}

\be
\delta \lambda = \frac{1}{2} \slashed\partial  \Phi \e - \frac{1}{24}
\slashed{H} \s_3 \e + \frac{1}{2} e^{\Phi}
\biggl[ \slashed{G}_{1} (i \s^2) + \frac{1}{2}
\left( \frac{1}{6}\slashed{G}_3  -m e^{-3A} \G^{123} \right)  \s^1 \biggr] \e\ ,
\label{dil1}
\ee
while the gravitino variations along $M_7$ and $S^3$, respectively, become
\ba
\label{susy1}
&& \delta \psi_{\mu} = D_{\mu} \e - \frac{1}{8} H_{\mu \nu \rho}
 \G^{\nu \rho} \s^3 \e - \frac{e^{\Phi}}{ 8} \biggl[ \slashed{G}_1 (i \s^2)
+ \left(  \frac{1}{6}\slashed{G}_3 -m e^{-3A} \G^{123}  \right) \s^1
\nn
&& \phantom{xxxx} +  \frac{e^{-3  A}}{2} \slashed{G}_{2} \G^{123} (i \s^2) \biggr]  \G_{\mu} \e
\ea
and
\ba
&& \delta \psi_{i}
= e^{-A} D^{S^3}_i\! \e - \frac{1}{2} \slashed \partial A   \G_i \e -  \frac{e^{\Phi}}{ 8}
\biggl[ \slashed{G}_1  (i \s^2)
+ \left(\frac{1}{6} \slashed{G}_3  -m e^{-3A} \G^{123} \right) \s^1
\nonumber\\
&& \phantom{xxxx} +  \frac{e^{-3  A}}{2}  \slashed{G}_{2} \G^{123}  (i \s^2) \biggr] \G_i \e\ .
\label{susy1a}
\ea
The indices $i=7,8,9$ denote the tangent space directions
of $S^3$ and $\mu =0,\dots, 6$ are $M_7$ indices. We have used that $D_i \e = D^{S^3}_i\! \e -\ha
e^A \slashed \partial A \G_i\e $, so that the covariant derivative $D^{S^3}_i\! $ is defined
entirely on the three-sphere as the notation indicates.

\no
We would like to rewrite the Killing spinor equations of massive IIA in terms of those
of type-IIB we just described. The key observation is the redefinition of the spinor (\ref{rotate}).
Plugging this into the gravitino variation in the $r$-direction and pulling through the
exponential factor $e^X$, one obtains
\ba
\label{psir}
&& \delta \psi_{r} =  e^{X} \biggl[  \frac{1}{2} \slashed \partial A \G_{r}
- \frac{e^{-A}}{4} \G^{\theta \phi} \s^3 + \frac{e^{\Phi}}{8} \biggl(m e^{-3 A}  \G^{\theta \phi}
 \G_r (i \s^2)
\nonumber\\
&&\phantom{xxxx}
 - \slashed{G}_1 (i \s^2) -  \frac{e^{-3 A}}{2} \slashed{G}_2 \G^{\theta \phi} \G_r \s^1
- \frac{1}{6} \slashed{G}_3 \s^1 \biggr)  \biggr]  \tilde{\e}\ ,
\ea
which is an algebraic condition that will lead to suitable projection conditions on the Killing spinor.
One can check that the two known solutions presented in \cite{Sfetsos:2010uq}
lead to such projections. For the T-dual of
$AdS_5 \times S^5$ one finds a single projection condition, while for the T-dual
of $AdS_3 \times S^3 \times T^4$, the vanishing of this
gravitino variation is equivalent to the imposition of two projection conditions.

\no
One can then use \eqn{psir} in the dilatino variation of massive IIA which then becomes
\ba
\label{dil2}
&&  \delta \lambda = e^{X} \left[ \frac{1}{2} \slashed \partial \Phi
  - \frac{1}{24}  \slashed{H} \s^3 \right] \tilde{\e} +
 \left[\frac{r^2 + 3 e^{4 A}}{r^2 + e^{4 A}} \G^{r}
 - \frac{2 r e^{2 A}}{r^2+ e^{4 A}} \G^{r \theta \phi} \s^3  \right] \delta \psi_r
\nonumber\\
 &&\phantom{xxxxxx} + e^{X} \frac{e^{\Phi}}{2} \left[ -\slashed{G}_1 \G^r (i \s^2) + \frac{1}{2}
 \left( m e^{-3 A}
 \G^{\theta \phi} (i \s^2)  - \frac{1}{6} \slashed{G}_3 \G^r \s^1\right) \right] \tilde{\e}\ .
\ea
The gravitino variation along $M_7$ may be similarly expressed as
\ba
\label{susy2}
&& \delta \psi_{\mu} = e^{X} \biggl[ D_{\mu}  - \frac{1}{8} H_{\mu \nu \rho}
\G^{\nu \rho} \s^3  + \frac{e^{\Phi}}{8} \biggl(e^{-3 A} m \G^{\theta \phi} (i \s^2)
\nn
&& \phantom{xxxxx}
- \slashed{G}_1 \G^{r}  (i \s^2)- \frac{e^{-3 A}}{2!} \slashed{G}_2 \G^{\theta \phi} \s^1
- \frac{1}{3!} \slashed{G}_3 \G^{r} \s^1 \biggr)  \G_{\mu} \biggr] \tilde{\e}\ .
\ea
Hence when $\delta \psi_r = 0$, as required by supersymmetry,
then the dilatino variation and the gravitino variations along $M_7$
resemble those of type-IIB. It is indeed easy to confirm that the Killing spinor equations
on $M_7$ are the same and that the Killing spinors are simply rotated
as described in \cite{Hassan:1999bv}. After incorporating the rotations involving $\e^{X}$,
one can simply redefine
\be
\tilde{\e}_{+} = \G^{r} \e_+\ , \qq \tilde{\e}_- =- \e_-\ , \qq \G^{r\theta \phi} = - \G^{123}\ ,
\label{324r}
\ee
so that, when $\delta \psi_r =0$, one maps (\ref{dil2}) to (\ref{dil1}) and (\ref{susy2}) to (\ref{susy1}).
The redefinition in $\e_{+}$ means that it flips chirality, in accordance with the
Killing spinors for type-IIB.

\no
The gravitino variation in the direction $\th$ can then be written as
\ba
&& \delta \psi_{\theta} = \frac{e^{2X}}{\sqrt{r^2 + e^{4 A}}} \left( e^{2A} \G^{r \phi} \s^3
 - \frac{e^{4A}}{r} \G^{r \theta} \right) \delta \psi_r
\nonumber\\
&& \phantom{xx}
+ \frac{e^{X} \sqrt{r^2+e^{4A}}}{r} \biggl[ e^{-A} \partial_{\theta} + \frac{1}{2} \G^{\theta \mu}
\partial_{\mu} A + \frac{e^{-A}}{4} \G^{r \phi} \s^3  +\frac{e^{\Phi}}{8}
\biggl( - m e^{-3 A} \G^{\phi} (i \s^2) \nonumber
\\
&& \phantom{xx}
 - \slashed{G}_1 \G^{r \theta} (i \s^2)
+\frac{e^{-3 A}}{2} \slashed{G}_2 \G^{\phi} \s^1 - \frac{1}{6} \slashed{G}_3 \G^{r \theta}
\s^1 \biggr) \biggr] \tilde{\e}\ ,
\ea
where of course the index in $\partial_\theta$ is curved, or simplifying further, as
\be
\delta \psi_{\theta} = e^{4 X} \G^{r \theta} \delta \psi_{r}
 +  \frac{e^{-A} \sqrt{r^2+e^{4A}}}{r} \left(  \partial_{\theta}
  + \frac{1}{2} \G^{r \phi} \s^3 \right) \tilde{\e}\ .
\ee
There is a similar variation for the remaining direction $\phi$ along $S^2$, so that
we may write compactly that
\be
\d\psi_\a = e^{2 X} \G_r \G_\a \d\psi_r +  \frac{e^{-A} \sqrt{r^2+e^{4A}}}{r} \left(D_\a
+ \ha \e_{\a\b} \G_r \G_\b \s_3\right)\tilde \e\ ,
\ee
where $\a,\b$ are tangent space indices along the unit 2-sphere directions $\th$ and $\phi$ of the T-dual
background and $\e_{\a\b}$ is the two-dimensional antisymmetric tensor with $\e_{\th\phi}=1$.
According to our normalizations $R_{\a\b} = g_{a\b}$. Hence,
one can readily check that the integrability condition of the $\d\psi_\a=0$
equation, arising from
$
[D_\m,D_\n]\e = {1\ov 4} R_{\m\n a b} \G^{ab}\e
$,
is indeed satisfied without a requirement for any extra projection condition.

\no
Hence upon satisfying $\d \psi_r =0$ in \eqn{psir}
we may map the massive IIA Killing spinor equations to those of type-IIB.
In fact, this equation becomes that of type-IIB along $S^3$ in \eqn{susy1a}
provided that in this theory the Killing spinors satisfy
\be
D^{S^3}_i\! \e = {1\ov 4} \G_{123}\G_i \e  \ ,
\label{kills3}
\ee
which is the Killing spinor equation on $S^3$. This is not a trivial statement in the sense that
there should be extra projections imposed on it in order to be fulfilled.
The solution to \eqn{kills3}
in the coordinate system \eqn{ds32} is readily found to be\footnote{We use the
frame
\be
e^\th = 2 d\th \ ,\qq e^\phi = 2 \cos\th d\phi\ ,\qq e^\psi = -2\sin \th d\psi\ ,
\ee
where the introduction of the minus sign is in accordance with \eqn{324r}.
}
\be
\e = \exp\left({\th \ov 2}\G_{\phi\psi} \right)  \exp\left(-{\phi+\psi \ov 2}\G_{\th\psi} \right)\e_0\ ,
\label{flaski1}
\ee
where $\e_0$ is a spinor that could depend on the $M_7$ coordinates.
Hence, given a solution of the type-IIB Killing spinor equations we should impose suitable projections
so that the solution eventually assumes the form \eqn{flaski1}. For example, for the case of the non-Abelian
T-dual to flat spacetime we easily see that by imposing the projection
\be
\G_x \e_0 = - \G_{\th\phi\psi}\e_0\ ,
\ee
into \eqn{flaski} we indeed obtain \eqn{flaski1}. In fact we may proceed further
and show that \eqn{flaski1} implies that the Lorentz--Lie (equivalently the Kosmann) derivative
on the Killing spinor of the original type--IIB theory vanishes.
We recall that the latter defines the action of
a vector on a spinor as \cite{Kosmann,FigueroaO'Farrill:1999va,Ortin:2002qb}
\be
\cL_{k} \e = k^\m D_\m\e + {1\ov 4} D_\m k_\n \G^{\m\n} \e\ .
\label{ksoom}
\ee
This derivation maps spinors to spinors and if $k^\m\del_\m$ is
a Killing vector then they obey the Lie-algebra
of the associated symmetry group.
In our case this symmetry algebra is generated by the left and right invariant vector fields
corresponding to the Maurer--Cartan forms with structure constants $\e_{abc}$ and $-\e_{abc}$, respectively.
Recall also that a Killing vector remains so in all conformally related metrics.
Since the original Killing spinor
corresponds to a background with $SU(2)_L\times SU(2)_R$ symmetry it transforms in
the direct product of the spinor representations of the two factors. The non-Abelian T-dual
background preserves just the $SU(2)_R$ factor. This is encoded in the
vanishing of the Kosmann derivative \eqn{ksoom}
for the right invariant Killing vectors.
After some algebraic manipulations one shows that demanding the latter condition implies \eqn{kills3},
thus fully justifying the use of the Kosmann derivative in the present context as introduced in
\cite{Sfetsos:2010uq}.

\no
Hence, our general conclusion is that, within our class of type-IIB
backgrounds \eqn{origbac}, \eqn{IIBfluxansatz} with
$SO(4)$ symmetry, a non-Abelian T-duality transformation with respect
to the $SU(2)$ subgroup giving rise to the
massive IIA background \eqn{NSsec} and \eqn{fliIIAmas}, reduces supersymmetry by half  if the 
Killing spinor transforms under the $SU(2)$ factor that we use to perform the T-duality transformation,
or leaves it intact if it does not transform at all.

\section{Massive IIA non-Abelian T-duals}
\label{sec:IIA}

For completeness we will also consider $SO(4)$ symmetric backgrounds in
massive IIA supergravity and their
non-Abelian duals with respect to an $SU(2)$ subgroup.
 We begin by establishing an $SO(4)$ invariant spacetime ansatz where we simply retain the singlets.
For the NS sector fields the metric is still given by (\ref{origbac}) as in the type-IIB case,
we omit the presence of a $B$-field along $S^3$ and the dilaton may depend only on coordinates of $M_7$.
The massive IIA ansatz for the fluxes is
\ba
&& F_0 = m \ ,
\nonumber\\
&& F_2 = G_2\ ,
\\
&& F_4 = G_4 + G_1\wedge {\rm Vol}(S^3)\ .
\label{formsIIA}
\nonumber
\ea
As for type-IIB one can consistently reduce the equations of motion of massive IIA on $S^3$ and obtain eventually
in the Einstein frame (\ref{EinactIIA}).
Using the redefinition (\ref{reddf}) one finds the following seven-dimensional action
\ba
\label{EinactIIAss}
&& \mathcal{L}_{\rm Einstein} = {R}
- 20 (\partial \Phi)^2 - 3 (\partial A)^2  - \frac{1}{12} e^{-8 \Phi} H^2
\nn
&& \phantom{xxxx} - \frac{1}{2} \left( m^2
e^{14 \Phi + 3 A} - 3 e^{4 \Phi - 2 A } + e^{10 \Phi  -3 A} G_1^2 +  \frac{1}{2} e^{6 \Phi + 3 A} G_2^2
+ \frac{1}{4!} e^{3 A - 2\Phi} G_4^2\right)
\nn
&& \phantom{xxxx} -  G_4 \wedge G_1 \wedge B + \frac{m}{3} G_1 \wedge B^3
+ \frac{1}{2} B^2 \wedge d C_1 \wedge G_1\ ,
\ea
where
\be
G_1 = dC_0 \ ,\qq G_2 = d C_1 + m B\ ,\qq G_4 = d C_3 - H \wedge C_1 + \frac{m}{2} B \wedge B\ .
\label{g1g2g5}
\ee
The RR fluxes transform according to \eqn{ppom} with the expressions for $P$ and $\hat P$ interchanged since
now the original background is in massive IIA and the final in type-IIB supergravity. Omitting the details,
which are given in Appendix B.2, we find that

\newpage
\ba
&&  \hat F_1 = - G_1 - m r dr \ ,
\nn
 && \hat F_3 =  e^{3A} \star_7 G_4 - r dr \wedge G_2
- \frac{r^3}{r^2 + e^{4 A}} G_1 \wedge \vol(S^2)
+ \frac{m  r^2 e^{4A} }{r^2 + e^{4 A}} dr \wedge \vol(S^2)\ ,
\\
&&  \hat F_5 = %(1 + *) \left[ \frac{e^{3 A} r^3}{r^2 + e^{4 A}} *_7 G_4
% + \frac{e^{4A} r^2}{r^2 + e^{4 A}} G_2 \wedge dr  \right] \wedge \vol(S^2) \ ,
{r^2 e^{3 A}\ov r^2 + e^{4A}}\left(r \star_7 G_4 + e^A dr\wedge G_2\right)\wedge \vol(S^2)
- e^{3A} \star_7 G_2- r dr\wedge G_4\ .
\nonumber
\ea
Finally, we note that an analysis along the lines of section 3
leads to the same conclusion that
supersymmetry is broken by half in this case as well.

\section{Explicit examples}
In this paper we have placed non-Abelian T-duality on a firmer footing.
Instead of being confined to near-horizon solutions
\cite{Sfetsos:2010uq,Lozano:2011kb},
 we are now in a position to generate
large families of solutions on the proviso that the original solution has an $S^3$ factor.
We can now simply match the original solution
 to our ansatz and read off the non-Abelian T-dual. We discuss below three such examples.

\subsection{PP-wave}

We begin by warming up with the maximally supersymmetric pp-wave in type-IIB supergravity \cite{Blau:2001ne}.
Here we have $SO(4) \times SO(4)$ isometry, so we have a choice of two three-spheres to T-dualise. Isolating the two $S^3$'s, the solution may be written as
\ba
ds^2 &=& 2 dx^{+} dx^{-} - \mu^2 \left[y^2 + z^2 \right] (dx^+)^2 + dy^2 + {1\ov 4} y^2ds^2(S^3)
 +  dz^2 + {1\ov 4} z^2ds^2(\tilde{S}^3)\ , \nn
F_5 &=&  \ha \mu y^3 dx^{+} \wedge dy \wedge \vol(S^3) + \ha  \mu z^3 dx^{+} \wedge dz \wedge \vol(\tilde{S}^3)\ ,
\ea
where again we use for the three-spheres the normalization $R_{ij}=\ha g_{ij}$.
We can now read off the field content in seven dimensions
\be
G_2 = \frac{1}{2} \mu y^3 dx^+ \wedge dy\ , \qq e^{A} = \frac{y}{2}\ .
\ee
We can then generate the T-dual solution of type-IIA
\ba
\label{IIBtoIIA}
ds^2 &=& 2 dx^{+} dx^{-} - \mu^2 \left[y^2 + z^2 \right] (dx^+)^2 + dy^2 + \frac{4}{y^2} dr^2
+ \frac{4 r^2 y^2}{16 r^2 + y^4} ds^2(S^2) \nn
&+& dz^2 + z^2 ds^2(\tilde{S}^3)\ ,
\nn
\Phi &=& - \frac{1}{2} \ln \left[ \frac{y^2}{4} \left( r^2 + \frac{y^4}{16} \right) \right] ,
 \quad B = \frac{16 r^3}{16 r^2 + y^4} \vol(S^2)\ ,
\\
F_2 &=& -\frac{1}{2} \mu y^3 dx^{+} \wedge dy\ , \quad F_4 = -\frac{8 r^3 \mu y^3}{16 r^2 + y^4}dx^{+}
 \wedge dy \wedge  \vol(S^2)\ .
\nonumber
\ea
It is also easy to see that supersymmetry is broken by one half.
Typically pp-waves of this form always preserve 16 supersymmetries in the kernel of $\G^{+}$.
 Plugging the solution into (\ref{psir}) one notes that these Killing spinors are subject to a projection condition $\G^{y r \theta \phi} \s^3 \tilde{\e}_{+} = \tilde{\e}_+$,
where we have used the subscript to denote Killing spinors satisfying $\G^{+} \tilde{\e}_+ = 0$.
The other sixteen Killing spinors not killed by $\G^{+}$ will also be subject to the same projector, so both standard Killing spinors
and supernumerary Killing spinors are cut by one half.
Indeed, this should not come as a surprise. In the process of constructing the non-Abelian T-dual we have deformed the original solution
so that the Ricci tensor has non-zero components other than $R_{++}$.

\no
We can now do the reverse transformation by reading off the transformation from section \ref{sec:IIA}. There is no need to rescale the fluxes as the fluxes
are now along $M_7$ only. The warp factor is then $e^{A} = \frac{z}{2}$ and the T-dual geometry in type-IIB,
after doing two non-Abelian T-dualities is
\ba
\label{IIAtoIIB}
ds^2 &=& 2 dx^{+} dx^{-} - \mu^2 \left[y^2 + z^2 \right] (dx^+)^2 + dy^2 + \frac{4}{y^2} dr^2 + \frac{4 r^2 y^2}{16 r^2 + y^4} ds^2(S^2) \nn
&+& dz^2 +  \frac{4}{z^2} d \tilde{r}^2 + \frac{4 \tilde{r}^2 z^2}{16 \tilde{r}^2 + z^4} ds^2(\tilde{S}^2)\ ,
\nn
e^{-2 \Phi} &=& \frac{(yz)^2}{16} \left( r^2 + \frac{y^4}{16} \right) \left( \tilde{r}^2 + \frac{z^4}{16} \right)\ ,
\nn  B &=& \frac{16 r^3}{16 r^2 + y^4} \vol(S^2) +  \frac{16 \tilde{r}^3}{16 \tilde{r}^2 + z^4} \vol(\tilde{S}^2)\ ,
\\
F_3 &=& \frac{z^3}{8} \star_7 F_4 - \tilde{r} F_2 \wedge d \tilde{r}\ ,
\nn
F_5 &=& (1+*) \left[ \frac{2 z^3 \tilde{r}^3}{16 \tilde{r}^2+z^{4}} \star_7 F_4
+ \frac{z^4 \tilde{r}^2}{16 \tilde{r}^2+z^4} F_2 \wedge d \tilde{r} \right]  \wedge \vol(\tilde{S}^2)\ ,
\nonumber
\ea
where $F_2$ and $F_4$ are given above in (\ref{IIBtoIIA}). One can check that the fluxes are of
the appropriate form so that we still have symmetry under the exchange $(r, y) \leftrightarrow (\tilde{r}, z)$.

\subsection{Type-IIB backgrounds with $SO(4) \times SO(4) \times  \mathbb{R}$ isometry}

In this section we consider the class of backgrounds constructed in \cite{LLM} that correspond to $\ha$-BPS states.
These contain two round three-spheres and a time-like Killing
vector. The metric is given by
\begin{equation}
ds^2=-h^{-2}(dt+V_idx^i)^2+h^2(dy^2+dx^i dx^i)+{1\ov 4}y\,e^G ds^2(S^3)
+{1\ov 4}y\,e^{-G}ds^2({\tilde S}^3)\ ,
\label{llmmet}
\end{equation}
where $i=1,2$ and $h, V_i$ and $G$ are functions of the $x^i$'s spanning an $\mathbb{R}^2$
and $y\geqslant 0$. They are related through the Killing spinor equations by
\begin{eqnarray}
&&h^{-2}=2\,y\cosh{G}\ , \qquad y\,\partial_y V_i=\epsilon_{ij}\partial_j z\ ,
\nonumber\\
&&y(\partial_i V_j-\partial_j V_i)=\epsilon_{ij}\partial_y z , \qquad z=\frac12 \tanh{G}\, .
\label{congfh3}
\end{eqnarray}
The non-trivial 5-form field strength is
\begin{equation}
F_5=G_2\wedge {\rm Vol}(S^3)+\tilde G_2 \wedge {\rm Vol}({\tilde S}^3)\ ,
\label{5formllm}
\end{equation}
where the two-forms $G_2$ and $\tilde G_2$ are along $t,x_i$ and $y$ and are given by
\begin{eqnarray}
&&8 G_2=dB_t\wedge (dt+V)+B_t dV+d{\hat B}\ ,
\nonumber\\
&&8 {\tilde G}_2=d{\tilde B}_t\wedge (dt+V)+{\tilde B}_t dV+d{\hat {\tilde B}}\ ,
\end{eqnarray}
with
\begin{eqnarray}
&&B_t=-\frac14 y^2 e^{2G}\, , \qquad {\tilde B}_t=-\frac14 y^2 e^{-2G}\, ,
\nonumber\\
&&d{\hat B}=-\frac14 y^3 \star_3 d\left(\frac{z+1/2}{y^2}\right)\, , \qquad
d{\hat {\tilde B}}=-\frac14 y^3 \star_3 d\left(\frac{z-1/2}{y^2}\right)\ \, ,
\end{eqnarray}
where the Hodge star is with respect to the three-dimensional metric with coordinates
the $x^i$'s and $y$.
The whole background can then be determined in terms of the function $z(x^1,x^2,y)$ satisfying
\begin{equation}
\label{eqz}
\partial_i \partial_i z+y\partial_y (\frac{\partial_y z}{y})=0\,
\end{equation}
and arising as the
integrability condition for the differential equations in \eqn{congfh3}. Solutions
to this are easily found by realizing that $\displaystyle \Phi ={z\ov y^2}$
satisfies the six-dimensional Laplace equation with $SO(4)$ rotational symmetry
and $y$ being the radial coordinate.

\no
An important issue in this class of backgrounds is regularity \cite{LLM}.
In order to avoid the singularity at $y=0$, the function $z(x^1,x^2,0)$
must only take the two possible values $\pm \frac12$ which are related to
the symmetry of the background under the
exchange of the two round 3-spheres.
The general solution of (\ref{eqz}) must then satisfy these boundary conditions in order to be regular.
Near $z=\frac12$, we have that $z=\ha - e^{-2 G}$ for $G$ large.
Simultaneously, as $y\to 0$, the function
$h^{-2}\simeq y e^G$ remains finite.
Then the part of the metric spanned by $y$ and the two 3-spheres behaves in this limit as
\begin{equation}
{1\ov 4} h^{-2}ds^2(S^3)+h^2 \left(dy^2+{y^2 \ov 4} ds^2({\tilde S}^3)\right)\ .
\end{equation}
One can also show that $V$ remains finite. For $z=-\frac12$ the same holds with
the two three-spheres interchanged.
The $\mathbb{R}^2$ plane has a natural interpretation as the phase space of one-dimensional
fermions in a harmonic potential \cite{LLM}. It is filled by quantum
Hall droplets where the fermions are localized. Their density $\r(x^1,x^2) = \ha - z(x^1,x^2,0)$
is a step function,
i.e. it takes the value $1$ inside the droplets and $0$ outside.

\no
The background \eqn{llmmet}, \eqn{5formllm} is of the general type \eqn{origbac}, \eqn{IIBfluxansatz}
with
\begin{equation}
e^A=\ha \sqrt{y}\,  e^{G/2}\ ,\qq  m=0\ ,\qq  G_1=0 \ ,\qq G_3=0\
\end{equation}
and non-vanishing $G_2$.
The metric of the non-Abelian T-dual solution is
\begin{eqnarray}
&&ds^2=-h^{-2}(dt+V_idx^i)^2+h^2(dy^2+dx^i dx^i)
\nonumber\\
&&\phantom{xxxx}  + 4 y^{-1}e^{-G}dr^2+  \frac{4r^2\, y\, e^G}{16 r^2+y^2e^{2G}}ds^2(S^2)
+  {1\ov 4} ye^{-G}ds^2(\tilde S^3)\, ,
\label{fjgkj1}
\ea
supported by
\ba
&&\Phi=-\frac12 \ln \left(y\, e^G (16 r^2+y^2 e^{2G})\right)\, ,
\qquad B=\frac{16 r^3}{16 r^2+y^2e^{2G}}{\rm Vol}(S^2)\ ,
\nonumber\\
&&F_2=-G_2\ , \qquad F_4=-\frac{16 r^3}{16 r^2+y^2e^{2G}}\,G_2\wedge {\rm Vol}(S^2)\ .
\nonumber
\end{eqnarray}
This background is singular at $y=0$, where the radii of both the two-and the three-spheres vanish,
unless $z=\frac12$. In this limit the part of the metric \eqn{fjgkj1} spanned by $y$, $r$ and the 2- and 3-spheres behaves as
\begin{eqnarray}
\label{hpart}
 h^2 \left[dy^2+4 dr^2+\frac{4 r^2}{1+16h^4r^2}ds^2(S^2)+{y^2\ov 4}  ds^2(\tilde S^3)\right]\ ,
\end{eqnarray}
which is non-singular. The same happens for the rest of NS-NS and RR fields. On the other hand
if $z=-\frac12$, $h^{-2}\simeq ye^{-G}$ and the metric and dilaton fields are singular.
Therefore the dual solution is non-singular only outside the droplets (in the original description)
where $z=\ha$.
This was expected since after the dualization
the symmetry under the exchange of the two 3-spheres is lost. In a similar fashion,
if we perform the non-Abelian
T-duality transformation on the $\tilde S^3$, we get a background that is singular
outside the droplets where
$z=\ha$ and non-singular inside them where $z=-\ha$.
Obviously, if we T-dualize with respect to both round three-spheres we obtain a
singular background everywhere in $\mathbb{R}^2$.
These general results are in agreement with the conclusions of \cite{Sfetsos:2010uq}, where the non-Abelian
T-dual of $AdS_5\times S^5$ with respect to the $SU(2)$ subgroup of the $SO(6)$ isometry group of
the five-sphere
was constructed.
Also we point out that there are interesting cases where the above regularity conditions are violated.
Notably,
if we use in place of the step function mentioned above, fermion distributions
at finite temperature \cite{Buchel:2004mc}
or in describing in the present context the so-called superstar solutions as in \cite{Caldarelli:2004mz}.

\no
Finally, given that the $SO(4)\times SO(4)$ symmetric massive deformation of ABJM considered in \cite{LLM,Bena:2004jw}
belongs as well to the general class of backgrounds, in this case of type-IIA, considered in this paper,
we can use non-Abelian T-duality to generate a type-IIB solution with non-vanishing $F_i$ for $i=1,3,5$, a NS $B$-field
and a metric with the same $y$, $r$, 2-sphere and 3-sphere components as in
(\ref{fjgkj1}).
We omit the details given the similarity with the present solution.

\subsection{Lifshitz symmetry solutions}

A particularly interesting class of geometries involves those exhibiting Lifshitz symmetry \cite{Kachru:2008yh}.
For concreteness, we will consider the solutions of \cite{Balasubramanian:2010uk} based on an $S^5$ internal geometry,
but will follow the notation of \cite{Donos:2010tu}.

\no
According to \cite{Donos:2010tu}, Lifshitz solutions with dynamical exponent $z=2$
in Einstein frame may be written as
\ba
ds^2 &=& r^2 \left[ 2 d x^+d  x^- + dx_1^2 + dx_2^2 \right] + \frac{dr^2}{r^2} + f (d x^+)^2 + ds^2(E_5)\ ,
\nn
F_5 &=& 4 (1+ *) \vol(E_5)\ ,
\\
G &=& d x^+ \wedge W\ , \quad P = g d x^+\ ,
\nonumber
\ea
where $W$, $f$ and $g$ satisfy
\be
\label{harm}
dx^+ \wedge d W = d *_E W = 0\ ,\qq - \nabla_E^2 f + 4 f = 4 |g|^2 +|W|^2_{E}\ .
\ee
Here $E_5$ denotes a compact Einstein manifold and we are using complex notation for the three-form and
the axion-dilaton\footnote{In terms of more usual string theory variables, these may be written as
$G = i e^{\frac{\Phi}{2}} \left( \tau d B - d C_{2}\right),$ $P = \frac{i}{2} e^{\Phi} d C_{0} + \frac{1}{2} d \Phi$,
where $\tau = C_{0} + i e^{-\Phi}$. }.

\no
The above requirement that $W$ be harmonic, (\ref{harm}), means that there are no solutions with non-zero $W$ for $S^5$,
but supersymmetric solutions \cite{Donos:2010tu} do exist for Sasaki-Einstein spaces such as $T^{1,1}$ \cite{Candelas:1989js} and $Y^{p,q}$ \cite{Gauntlett:2004yd}
which are topologically $S^2 \times S^3$. However, in spite of these spaces having the correct topology, neither possess a round $S^3$ fitting into our ansatz,
so we confine ourselves to $W=0$ with $E_5$ being $S^5$. To recover the
solution of \cite{Balasubramanian:2010uk} one simply takes $W=0$ with $f$ and $g$ only depending on the coordinate $x^+$.
A further subclass considers the case where $f$ is a constant, which following \cite{Donos:2010tu},
we also take to be the identity. In this case from (\ref{harm}) we have $g = e^{i \beta}$, where $\beta \in [0,\pi/2]$.
Then introducing the usual fibration of $S^5$,
\be
ds^2 = d \tilde{\theta}^2 + \sin^2 \tilde{\theta} d \tilde{\phi}^2 + \cos^2 \tilde{\theta} ds^2(S^3),
\ee
we can perform a non-Abelian T-duality to get the type-IIA solution.\footnote{Note that in this case the $S^3$ metric is normalized such that $R_{ij}=2g_{ij}$.}
After transforming to the string frame, one can simply read off the T-dual solution
from the formulae in section \ref{sec:IIBduals}, giving the following expressions
\ba
e^{\Phi} &=& e^{2 \cos \beta x^+}\ , \nn
e^{A} &=& \frac{1}{2} e^{\frac{1}{2} \cos \beta x^+} \cos \tilde{\theta}\ , \nn
G_1 &=& 2 \sin \beta e^{-2 \cos \beta x^+} d x^+ \ ,
\\
G_2 &=& \frac{1}{2}  \cos^3 \tilde{\theta} \sin \tilde{\theta} d \tilde{\theta} \wedge d \tilde{\phi}\ ,
\nonumber
\ea
with $G_3=m=0$ and
where we have introduced tildes to differentiate angles on $M_7$ from angles on $S^3$.

\no
One final interesting point pertains to supersymmetry.
 As is discussed in \cite{Donos:2010tu}, these solutions prior to T-duality generically
preserve two supersymmetries which are further enhanced to eight supersymmetries when $E_5 = S^5$ \cite{Das:2006dz}.
These eight supersymmetries are those preserved by
%\be
$\G^{+} \e = 0, ~ \G^{+- 1 2} \e = i \e$,
%\ee
where we have used complex spinors $\e = \e_1 + i \e_2$. The non-Abelian solution is subject to the additional
projection
%\be
%\left[ \sin \theta \G_{r}^{~\tilde{\theta}} - \G^{\theta \phi} \s^3
%+ \cos \theta \G^{\tilde{\theta} \tilde{\phi}} \G^{r \theta \phi} \s^1 \right] \tilde{\e} = 0\ ,
%\ee
found by imposing \eqn{psir} and therefore it preserves four supercharges.

\section{Concluding remarks}

In this paper we have put non-Abelian duality on a firmer footing by showing that it relates backgrounds that give rise to the same consistent truncation in
seven dimensions.
An important drawback of the original $\sigma$-model derivation in \cite{delaossa:1992vc} was the fact that
it was not possible to reconstruct the original background from the dual one due to its lack of isometries.
This problem is sorted out in the supergravity formulation, where any solution of the seven-dimensional theory can be uplifted to
both type-IIB and massive IIA supergravities. As we have seen, provided the original type-IIB solution has an RR $F_3$-flux along $S^3$,
a solution to massive IIA results. This provides quite a general set-up in which to generate solutions to massive type-IIA supergravity.

\no
We have also made a step in understanding supersymmetry breaking under non-Abelian T-duality. Through a mapping of the Killing
spinor equations and an expected redefinition of the Killing spinors, we have seen that the supersymmetry conditions get
mapped modulo an additional consistent condition that can break half the supersymmetry.

\no
Since a $B$-field along the $SU(2)$ directions of the dual background is also generated, non-Abelian duality
can be useful as well as a way to produce solutions with a  non-vanishing $B$. This raises an open question about D-brane probe
dynamics \cite{Burgess:2003mm} and the role of the $B$-field in overcoming natural repulsion between probe branes (see \cite{Lee:2008ha}).
Related to D-branes and their T-duals is a very pertinent question regarding the charges of the T-dual geometry. Since the resulting geometry
is non-compact, some form of regularization of the flux integrals will be required. On top of this, we can ask if there is an AdS/CFT
picture and if the large N limit of the solutions could be useful in the spirit of \cite{Maldacena:1999mh} to describe non-commutative gauge theories.
Furthermore, if a dual CFT picture can be understood, since supersymmetry is broken by a half, it would certainly be interesting to understand
this from the CFT point of view.

\no
As we have already mentioned, we expect that more general non-Abelian duality transformations based on larger symmetry groups
or acting with fixed points will also fit into this picture. The most straightforward extension is the construction
of non-Abelian $SU(2)$ duals in which the $SU(2)$ acts with fixed points. In this case, and based on the examples presented in \cite{Lozano:2011kb},
we conjecture that the original and dual backgrounds will also give rise to consistent truncations to lower-dimensional theories.
On that, we note that as our reduction on $S^3$ from type-IIB does not give rise to the expected maximal supergravity in
seven dimensions \cite{Samtleben:2005bp},
it would be satisfactory if the full reduction could be identified, or if the origin of the massive multiplets could be elucidated.

\no
We have seen in subsection 2.3 that in order to reproduce \eqn{D7action} from the action 
of massive IIA on the T-dual background with $\mathbb{R}\times S^2$ topology, 
we have to take a correlated limit in which the coupling constant of the ten-dimensional theory is taken 
infinitely large. 
As already mentioned in the main text, in analogy with the exact CFT models investigated in \cite{Polychronakos:2010hd} 
this implies that the non-Abelian T-dual background effectively captures states corresponding 
to group theory representations with infinitely high highest weight.
This also resonates with a result of \cite{Sfetsos:2010uq} in which the non-Abelian T-dual of an $SU(2) \subset SO(6)$ 
of the $AdS_5 \times S^5$ background gave rise to a solution whose M-theory lift captures generic features of the geometries
proposed in \cite{Gaiotto:2009gz} as gravity duals to $N = 2$ gauge theories. 
These features correspond to a zooming of the generic geometries
presumably associated to high spin states in the dual CFT. It is important to pursue work that 
substantiates further this idea.

\no
Finally, it would be interesting to examine the effect of non-Abelian T-duality on supergravity solutions with
interesting four-dimensional cosmological interpretations. Since T-duality breaks the $SO(4)$ symmetry of solutions with
homogeneity and isotropy down to $SO(3)$, it implies that the homogeneity is lost. The important physical question in the present context 
is to investigate the Big-Bang scenario and the fate of the initial singularity.

\subsection*{Acknowledgements}

\no
We are grateful to Bert Janssen, Hai Lin, Hong Lu, Patrick Meessen, Tom\'as Ort\'{\i}n and Henning Samtleben
for useful discussions.
The research of G. Itsios has been co-financed by
the ESF
and Greek national funds through the Operational Program "Education and Lifelong Learning" of the NSRF - Research Funding Program: "Heracleitus II.
 Investing in knowledge in society through
the European Social Fund''. He would also like to thank the CFP of University of Porto
for hospitality within the framework of the LLP/Erasmus Placements 2011-2012.
Y. Lozano and E. \'O Colg\'ain have been partially supported by the research grants MICINN-09-FPA2009-07122 and MEC-DGI-CSD2007-00042.

\appendix

\section{Review of type-II supergravities}

In this appendix we review aspects of type-II supergravities
relevant to this work.

\subsection{Type-IIB supergravity}

The action of type-IIB supergravity is given by
\ba
&& S_{\rm IIB} = {1\ov 2\kappa^2}\int_{M_{10}}\sqrt{-g} \Bigg[e^{-2\Phi}
\left(R + 4 (\del\Phi)^2 -{H^2\ov 12}\right)   -\ha\left(F_1^2 +{F_3^2\ov 3!}
+\ha {F_5^2\ov  5!}\right)\Bigg]
\nonumber\\
&& \phantom{xxxxxxx}
- \ha C_4 \wedge H \wedge dC_2\ ,
\label{actIIB}
\ea
where the field strengths in terms of the potentials are
\be
H= dB\ ,\qq  F_1 = dC_0\ ,\qq F_3 = dC_2 - C_0 H\ ,\qq F_5 = dC_4 - H \wedge C_2\ .
\ee
In addition, $F_5$ has to be self-dual.
The Bianchi identities are
\be
\label{BianchiIIB}
dH= 0\ ,\qq dF_1= 0\ ,\qq dF_3 = H\wedge F_1\ ,\qq dF_5 = H\wedge F_3\ .
\ee
Einstein's equations that follow from varying the metric are
\ba
&& R_{\m\n}+2 D_\m D_\n\Phi - {1\ov 4} H^2_{\m\n}
%=e^{2\Phi}\left[
%\ha \left((F_2^2)_{\m\n} -{1\ov 4} g_{\m\n}F_2^2\right)
 %+ {1\ov 12 } \left((F_4^2)_{\m\n} -{1\ov 8} g_{\m\n}F_4^2\right)
%- {m^2\ov 4}\right]\ ,
\nonumber\\
&& \phantom{xxxxxx}
=e^{2\Phi}\left[\ha (F_1^2)_{\m\n} + {1\ov 4} (F_3^2)_{\m\n} + {1\ov 96} (F_5)^2_{\m\n}
- {1\ov 4} g_{\m\n}
\left( F_1^2 +{1\ov 6} F_3^2 \right)\right]\ .
\label{EinIIB}
\ea
Note the fact that $F_5^2 =0$ due to the self-duality condition $\star F_5 = F_5$. The equation coming from varying the dilaton is
\be
R + 4 D^2 \Phi - 4 (\del \Phi)^2 - {1\ov 12} H^2 = 0\ .
\label{dilaeq}
\ee
Finally, from the variation of the various fluxes we obtain
\ba
&& d\left(e^{-2\Phi}\star H\right) - F_1\wedge \star F_3 - F_3\wedge F_5 = 0\ ,
\nonumber\\
&& d \star F_1 + H\wedge \star F_3 =0\ ,
\nonumber\\
&&
 d \star F_3 + H\wedge  F_5 =0\ ,
\label{fluxIIB}
\\
 && d \star F_5 - H\wedge  F_3 =0\ .
\nonumber
\ea
The equation of motion for $F_5$ is equivalent to the Bianchi identity for the $5$-form,
as it is self-dual.

\subsection{Massive IIA supergravity}

Having introduced type-IIB we now turn to massive type-IIA supergravity.
In making the transition from type-IIA supergravity to massive IIA, one simply
modifies the definitions of the field strengths by introducing a mass parameter $m$
as
\be
H= dB\ ,\qq  F_2 = dC_1 + m B \ ,\qq F_4 = dC_3 - H\wedge C_1 + {m\ov 2}B\wedge B \ .
\ee
The relative coefficients are fixed so that the field strengths are invariant
under the gauge transformations
\be
\d B= d\L\ ,\qq \d C_1 = -m \L\ ,\qq \d C_3 = - m \L \wedge B\ ,
\label{gautr}
\ee
where $\L$ is a one-form.
The Bianchi identities become
\be
\label{BianchiIIA}
dH= 0 \ ,\qq dF_2= m H\  ,\qq dF_4 = H\wedge F_2\ ,
\ee
which means that $m$ is like an expectation value for an $F_0$ term.
The action is the same as in type-IIA theory, but with the new definitions for the field
strengths. The topological term can therefore be written as
\be
-{1\ov 2} \int_{M_{11}} F_4 \wedge F_4 \wedge H = \cdots =  -{1\ov 2} \int_{M_{10}}
dC_3 \wedge dC_3 \wedge B + {m\ov 3} dC_3 \wedge B^3 + {m^2 \ov 20} B^5\ ,
\ee
using an obvious notation for the powers of the forms.

\no
Hence, the action of the massive IIA  supergravity is
\ba
&& S_{\rm Massive\ IIA}  =  {1\ov 2\kappa^2}\int_{M_{10}} \sqrt{-g} \Bigg[ e^{-2\Phi}
\left(R + 4 (\del\Phi)^2 -{H^2\ov 12}\right)  -\ha\left(m^2 + {F_2^2\ov 2} +{F_4^2\ov 4!}\right)\Bigg]
\nonumber\\
&&\phantom{xxxxxxxxxx} - {1\ov 2}\left( dC_3 \wedge dC_3 \wedge B
+  {m\ov 3} dC_3 \wedge B^3 + {m^2 \ov 20} B^5\right)\ .
\label{act2}
\ea
Einstein's equations are
\ba
&& R_{\m\n}+2 D_\m D_\n\Phi - {1\ov 4} H^2_{\m\n}
%=e^{2\Phi}\left[
%\ha \left((F_2^2)_{\m\n} -{1\ov 4} g_{\m\n}F_2^2\right)
%+ {1\ov 12 } \left((F_4^2)_{\m\n} -{1\ov 8} g_{\m\n}F_4^2\right)
%- {m^2\ov 4}\right]\ ,
\nonumber\\
&& \phantom{xxxxxx}
=e^{2\Phi}\left[\ha (F_2^2)_{\m\n} + {1\ov 12} (F_4^2)_{\m\n}- {1\ov 4} g_{\m\n}
\left(\ha F_2^2 +{1\ov 24} F_4^2 + m^2\right)\right]\ .
\label{Einsteq}
\ea
The flux equations are
\ba
&& d\left(e^{-2\Phi}\star H\right) - F_2\wedge \star F_4 -  {1\ov 2} F_4\wedge F_4 = m \star F_2\ ,
\nonumber\\
&& d \star F_2 + H\wedge \star F_4 =0\ ,
\label{fluxx}
\\
&&
 d \star F_4 +  H\wedge  F_4 =0\ .
\nonumber
\ea
This set of equations is consistent with the Bianchi identities as it can be seen by applying
to each one of them the exterior derivative. The dilaton equation (\ref{dilaeq}) is the same as before.

\subsection{Supersymmetry}

\no
Our conventions for supersymmetry variations follow those of \cite{Hassan:1999bv}.
To package these variations we find it handy to introduce
a Killing spinor comprising of real Majorana--Weyl spinors
\be
\e = \left( \begin{array}{c} \e_+ \\ \e_- \end{array}\right).
\ee
In type-IIB we have $\G^{11} \e = \e$, while in type-IIA the conventions
are such that $\G^{11} \e_{\pm} = \mp \e_{\pm}$.
Using Pauli matrices, the type-IIA Killing spinor equations can be written as
\ba
&& \delta \lambda = \frac{1}{2} \slashed \del \Phi \e - \frac{1}{24} \slashed{H} \s_3 \e
+ \frac{1}{8} e^{\Phi} \biggl[  5 m \s^1 + {3\ov 2} \slashed{F}_2  (i \s^2)
+ {1\ov 24} \slashed{F}_4 \s^1 \biggr] \e\ ,
\nonumber\\
&& \delta \psi_{\mu} = D_{\mu} \e - \frac{1}{8} H_{\mu \nu \rho} \G^{\nu \rho} \s^3 \e
+ \frac{e^{\Phi}}{ 8} \biggl[ m  \s^1
+ {1\ov 2} \slashed{F}_2 (i \s^2) + {1\ov 24} \slashed{F}_{4} \s^1 \biggr]  \G_{\mu} \e\ ,
\label{killIIA}
\ea
where $\displaystyle D_\m \e = \del_\m \e + {1\ov 4} \om_\m^{ab} \G_{ab} \e $.
The Killing spinor equations of type-IIB are
\ba
&& \delta \lambda = \frac{1}{2} \slashed \del \Phi \e - \frac{1}{24} \slashed{H} \s_3 \e
+ \frac{1}{2} e^{\Phi} \biggl[ \slashed{F}_{1} (i \s^2) + \frac{1}{12}\slashed{F}_3  \s^1 \biggr] \e\ ,
\nonumber
\\
&& \delta \psi_{\mu} = D_{\mu} \e - \frac{1}{8} H_{\mu \nu \rho} \G^{\nu \rho} \s^3 \e
 - \frac{e^{\Phi}}{ 8} \biggl[ \slashed{F}_1 (i \s^2)
+ {1\ov 6} \slashed{F}_3 \s^1 +  \frac{1}{240} \slashed{F}_{5} (i \s^2) \biggr]  \G_{\mu}\e\ ,
\label{killIIB}
\ea
where as always we are using the notation $\displaystyle \slashed{F}_n \equiv  F_{i_1\dots i_n} \G^{i_1 \dots i_n}$.

\section{Details of various KK reductions}

\subsection{Reduction of type-IIB on $S^3$}

\no
In performing the reduction on $S^3$ we first note that the type-IIB
Bianchi identities \eqn{IIBfluxansatz} imply that
\ba
&& dG_1= 0 \ ,\qq dG_3 = H \wedge G_1\ ,
\nonumber\\
&& dG_2 = -m H \ , \qq d(e^{-3 A}\star_7 G_2) + H\wedge G_3 = 0 \ .
\label{biaiIIB}
\ea
The first three relations may be integrated to give the field content \eqn{fielstreIIB}.
Similarly the type-IIB flux equations of motion \eqn{fluxIIB} imply that
\ba
&& d(e^{3 A-2 \Phi}\star_7 H )- e^{3A} G_1 \wedge \star_7 G_3 - G_3\wedge G_2
+m  e^{-3A} \star_7 G_2= 0\ ,
\nonumber\\
&& d(e^{3 A} \star_7 G_1) + e^{3 A} H\wedge\star_7 G_3 =0 \ ,
\label{fluxIIA}
\\
&& d(e^{3 A}\star_7 G_3)+ H\wedge G_2\ =0 \ .
\nonumber
\ea
We are after an effective seven-dimensional action which can capture this reduction procedure.
Variations of this action with respect to these potentials will give
equations of the form $d(\dots \star_7 G_{1,2,3})=\cdots $.
This implies that the last of (\ref{biaiIIB})
will arise from this action upon varying with respect to $C_1$.

\no
The Einstein equations \eqn{EinIIB} reduce to a single constraint equation along $S^3$
\ba
\label{Eincons1}
{ e^{-2 A}\ov 2} -  3 (\del A)^2   - D^2 A
+ 2 \del A\cdot  \del \Phi
=  {e^{2 \Phi}\ov 4} \biggl[ m^2 e^{-6 A}
 - G_1^2 + \ha e^{-6 A} G_2^2  - \frac{1}{6} G^2_3 \biggr]\ ,
\ea
where we have adopted the same normalisation as
\cite{Sfetsos:2010uq}, namely $R_{mn} = \frac{1}{2} g_{mn}$,
and the seven dimensional Einstein equation\footnote{One should make use of the
identity $\displaystyle \frac{1}{96} (\star_7 G_2)^2_{\mu \nu}
=   \frac{1}{4} (G_2)^2_{\mu \nu} -\frac{1}{8} g_{\mu \nu} G_2^2
$.}
%\ba
%R_{\mu \nu} &=& - 2 \nabla_{\nu} \nabla_{\mu} \Phi + 3 \nabla_{\nu} \nabla_{\mu} A + 3 \partial_{\nu} A \partial_{\mu} A   + \frac{1}{4} (H)^2_{\mu \nu}
%\nn &+& e^{2 \Phi} \biggl[ \frac{1}{2} (G_1)^2_{\mu \nu} + \frac{1}{4} (G_3)^2_{\mu \nu} + \frac{1}{4} e^{-6 A}  (G_2)^2_{\mu \nu} \nn &+& \frac{1}{96} e^{-6A} (*_7 G_2)^2_{\mu \nu} - \frac{1}{4} g_{\mu \nu} \left( m^2 e^{-6A} + G_1^2 + \frac{1}{6} G_3^2 \right) \biggr].
%\ea
%Note that all fields appearing here are seven-dimensional, including the metric $g_{\mu \nu}$. We can finally use the identity
%\be
%\frac{1}{96} (*_7 G_2)^2_{\mu \nu} = -\frac{1}{4(2!)} g_{\mu \nu} G_2^2 + \frac{1}{4} (G_2)^2_{\mu \nu},
%\ee
%to rewrite this in the more usual fashion
\ba
\label{D7Ein}
&& R_{\mu \nu} = - 2 D_{\mu} D_{\nu} \Phi+  3 D_{\mu} D_{\nu} A
+ 3 D_{\mu} A D_{\nu} A +  \frac{1}{4} (H)^2_{\mu \nu}
\nn
&& \phantom{xxxx}
+ \ha  e^{2 \Phi}\biggl[(G_1)^2_{\mu \nu}  +  e^{-6 A}  (G_2)^2_{\mu \nu}
 + \frac{1}{2} (G_3)^2_{\mu \nu}
\\
&& \phantom{xxxx} -  \frac{1}{2} g_{\mu \nu} \left( m^2 e^{-6A}
+ G_1^2 + \frac{1}{2} e^{-6 A} G_2^2 + \frac{1}{3!} G_3^2 \right) \biggr]\ .
\nonumber
\ea
The one remaining equation to be considered is the type-II
supergravity dilaton equation (\ref{dilaeq}). In terms of seven-dimensional fields it reads
\be
\label{ricciC}
R - 6 D^2 A - 12 (\partial A)^2 + \frac{3}{2} e^{-2A} + 4 D^2 \Phi
+ 12 \del A \cdot \del \Phi -4 (\partial \Phi)^2 -\frac{1}{12} H^2= 0\ .
\ee
One can repackage these equations of motion in an action with Lagrangian density
\ba
\mathcal{L} &=& e^{3 A-2 \Phi} \left( R + 6 (\del A)^2
+ 4 (\del \Phi)^2 - 12 \del A\cdot \del\Phi
 - \frac{1}{12} H^2 \right)
\nonumber\\
&-& \frac{1}{2} \left(m^2 e^{-3A} - 3 e^{A-2 \Phi} +  e^{3 A} G_1^2
+ \frac{e^{-3A}}{2} G_2^2 + \frac{e^{3A}}{6} G_3^2 \right)
\label{kgk22}
\\
&+& G_2 \wedge C_2 \wedge H\ ,
\nonumber
\ea
where the last line is a topological term.
In deriving the expression for the action we have made use of (\ref{ricciC}).
As a quick consistency check, one can confirm that the constraint
 equation from the
Einstein equations (\ref{Eincons1}) appears by varying the action with
respect to the scalar $A$, while (\ref{ricciC})
appears from varying the dilaton. Naturally, (\ref{D7Ein}) appears from
varying the action with respect to the metric.

\no
We can now perform the conformal transformation
\be
\label{rescale}
g_{\mu \nu} = e^{\frac{4 \Phi - 6  A}{5} } \hat{g}_{\mu \nu}\ ,
\ee
resulting in the Einstein frame action with Lagrangian density
\ba
&& \mathcal{L}_{\rm Einstein} = {R}  - \frac{24}{5} (\partial A)^2
- \frac{4}{5} (\partial \Phi)^2 +  \frac{12}{5} \partial A \cdot \partial \Phi
- \frac{1}{12} e^{\frac{12}{5} A - \frac{8}{5} \Phi} H^2
\nn
&& \phantom{xxx} - \frac{1}{2} \left( m^2 e^{- \frac{36}{5} A + \frac{14}{5} \Phi} - 3 e^{-\frac{16}{5} A
+ \frac{4}{5} \Phi } + e^{2 \Phi} G_1^2 +  \frac{1}{2} e^{-\frac{24}{5} A + \frac{6}{5} \Phi} G_2^2 + \frac{1}{3!} e^{\frac{12}{5} A
+ \frac{2}{5} \Phi} G_3^2\right)
\nn
&& \phantom{xxx}  +  G_2 \wedge C_2 \wedge H\ ,
\label{EEEUN}
\ea
where the last line is a topological term and where we have dropped the hats.

\subsection{Reduction of massive IIA on the non-Abelian T-dual}

Here we present some details on the steps involved in this reduction.
The metric (\ref{NSsec}) is  of the form
\be
 ds^{2}_{10}=ds^{2}(M_{7})+f_{1}^{2}dr^{2}+f_{2}^{2}ds^{2}(S^{2})\ ,
\ee
where $f_1$ and $f_2$ depend on the coordinates of $M_7$ and in addition $f_2$ depends on $r$ as well.
The result for the Ricci scalar is
\ba
&& \hat{R}  =    R-2\nabla^{2}\ln f_{1}-4\nabla^{2}\ln f_{2}
-2(\partial\ln f_{1})^{2}-6(\partial\ln f_{2})^{2}-4(\partial\ln f_{1})\cdot(\partial\ln f_{2})
\nonumber\\
          && \phantom{xxx} -\frac{2}{f_{1}^{2}}\left\{2\partial^{2}_{r}\ln f_{2}
+3(\partial_{r}\ln f_{2})^{2}\right\}+\frac{2}{f_{2}^2}\ .
\ea
If we substitute the expressions for $f_{1},f_{2}$ as read off from (\ref{NSsec}), we obtain
\ba
&&  \hat{R}  =R+\frac{4r^{2}e^{4A}+6e^{8A} -2r^{4}}{(r^2+e^{4A})^{2}}D^2 A
+\frac{40r^{2}e^{4A}-12e^{8A}-4r^{4}}{(r^{2}+e^{4A})^{2}}(\partial A)^{2}
\nonumber\\
         & &\phantom{xxx} +\frac{2e^{-2A}}{(r^{2}+e^{4A})^{2}}(r^{4}+3e^{4A}r^{2}+9e^{8A})\ .
\ea
The corresponding expressions for the dilaton and the NS three-form are
\be
(\partial\hat{\Phi})^{2} = (\partial\Phi)^{2}+
\frac{(r^{2}+3e^{4A})^{2}}{(r^{2}+e^{4A})^{2}}(\partial A)^{2}
-2\frac{r^{2}+3e^{4A}}{r^{2}+e^{4A}}\partial A\cdot \partial\Phi+\frac{r^{2}}{(r^{2}+e^{4A})^{2}}\
\ee
and
\be
\frac{\hat{H}^{2}}{12}=\frac{H^{2}}{12}+\frac{8r^{2}e^{4A}}{(r^{2}
+e^{4A})^{2}}(\partial A)^{2}+\frac{e^{-2A}}{2}\frac{(r^{2}+3e^{4A})^{2}}{(r^{2}+e^{4A})^{2}}\ .
 \ee
The RR flux terms are
\ba
&& \frac{1}{2}\hat{F}_{2}^{2}=m^{2}r^{2}e^{-4A}+r^{2}e^{2A}G_{1}^{2}+\frac{1}{2}G_{2}^{2}\ ,
\nonumber\\
&&\frac{1}{4!}\hat{F}_{4}^{2}
=e^{6A}G_{1}^{2}+\frac{r^{2}e^{-4A}}{2!}G_{2}^{2}
+\frac{r^{2}e^{2A}}{3!}G_{3}^{2}+\frac{e^{6A}}{4!}(\star_{7}G_{3})^{2}\ .
 \ea
We also note the identity $(\star_{7}G_{3})^{2}=-4 G_{3}^{2}$.
Finally, the topological term of the massive IIA theory becomes
\be
 -\frac{1}{2}\int\limits_{M_{11}}\hat{F}_{4}\wedge\hat{F}_{4}\wedge\hat{H}
= \int\limits_{M_{10}}\frac{r^4}{r^{2}+e^{4A}}C_{2}\wedge G_{2}\wedge H\wedge dr\wedge \vol(S^{2})\ .
\ee

\subsection{Reduction of massive IIA on $S^3$}

The Bianchi identities (\ref{BianchiIIA}) of massive IIA supergravity imply that
\be
dG_1= 0 \ ,\qq dG_2 = m H \ ,\qq dG_4 = H \wedge G_2\ ,
\ee
which can be integrated to give the field content (\ref{g1g2g5}).
Similarly the massive IIA flux eqs. (\ref{fluxx}) imply that
\ba
&& d(e^{3 A-2 \Phi}\star_7 H )- e^{3A} G_2 \wedge \star_7 G_4 - G_1\wedge G_4 = m e^{3A} \star_7 G_2\ ,
\nonumber\\
&& d(e^{3 A} \star_7 G_2) + e^{3 A} H\wedge\star_7 G_4 =0 \ ,
\label{fluxIIB2}
\\
&& d(e^{3 A}\star_7 G_4)+ H\wedge G_1 =0 \ ,\qq d(e^{-3 A}\star_7 G_1)+ H\wedge G_4 =0\ .
\nonumber
\ea
The Einstein equations reduce to
\be
  \frac{1}{2}e^{-2A}-3(\partial A)^{2}+2\partial A\cdot\partial\Phi-D^{2}A
= -\frac{e^{2\Phi}}{4}\left(m^{2}-e^{-6A}G_{1}^{2}+\frac{1}{2}G_{2}^{2}+\frac{1}{4!}G_{4}^{2}\right)
\ee
and the seven-dimensional Einstein equations
\ba
&& R_{\mu\nu}-3\partial_{\mu}A\partial_{\nu}A-3D_{\mu}D_{\nu}A+2D_{\mu}D_{\nu}\Phi
-\frac{1}{4}(H^{2})_{\mu\nu}
\nonumber\\
&& \phantom{xxxxxx} = \frac{e^{2\Phi - 3 A}}{2}
\Bigg\{e^{-3A}G_{1\mu}G_{1\nu}+e^{3A}(G_{2}^{2})_{\mu\nu}
+\frac{e^{3A}}{3!}(G_{4}^{2})_{\mu\nu}
\\
&&\phantom{xxxxxx}  -\frac{g_{\mu\nu}}{2}
\left(e^{3A}m^{2}+e^{-3A}G_{1}^{2}+\frac{e^{3A}}{2}G_{2}^{2}+\frac{e^{3A}}{4!}G_{4}^{2}\right) \Bigg\}\ .
\nonumber
\ea
The dilaton equation is the same as in \eqn{ricciC}.

\no
The equations above can be obtained from an action with Lagrangian density
\ba
&& \mathcal{L} =
e^{3A-2\Phi}\left(R + 6 (\del A)^2 - 12\del \Phi\cdot \del A
+ 4(\del\Phi)^2-\frac{H^2}{12} +\frac{3}{2} e^{-2A} \right)
\nonumber\\
&&
\phantom{xxx}
-\frac{1}{2}\left(m^2 e^{3A} + e^{-3A}G_{1}^{2}+\frac{e^{3A}}{2}G_{2}^{2}+\frac{e^{3A}}{4!}G_{4}^{2}\right)
\\
&&
\phantom{xxx}
-G_{4}\wedge G_{1}\wedge B+\frac{m}{3}G_{1}\wedge B^{3}
+ \frac{1}{2}B^{2}\wedge dC_{1}\wedge G_{1}\ ,
\nonumber
\ea
where the last line is a topological term.
Passing to the Einstein frame using \eqn{rescale} we obtain
\ba
\label{EinactIIA}
&& \mathcal{L}_{\rm Einstein} = {R}
- \frac{24}{5} (\partial A)^2 - \frac{4}{5} (\partial \Phi)^2
 +  \frac{12}{5} \partial A \cdot \partial \Phi
- \frac{1}{12} e^{\frac{12}{5} A - \frac{8}{5} \Phi} H^2
\nn
&& \phantom{x}
- \frac{1}{2} \left( m^2 e^{\frac{14}{5} \Phi - \frac{6}{5} A }
 - 3 e^{ \frac{4}{5} \Phi -\frac{16}{5} A  }
+ e^{-6 A + 2 \Phi} G_1^2 +  \frac{1}{2} e^{\frac{6}{5} A + \frac{6}{5} \Phi} G_2^2
+ \frac{1}{4!} e^{\frac{18}{5} A -\frac{2}{5} \Phi} G_4^2\right)
\nonumber\\
&& \phantom{x}  - G_4 \wedge G_1 \wedge B + \frac{m}{3} G_1 \wedge B^3
+ \frac{1}{2} B^2 \wedge d C_1 \wedge G_1\ .
\ea

\section{Details of derivation of the T-dual Ramond fields}

In computing the non-Abelian dual Ramond fields we need the Hodge
duals as well, since in the definition of the bispinors \eqn{bisp1} and \eqn{bisp2}
we use the democratic formulation in which all forms of degree up to
ten appear \cite{Bergshoeff:2001pv}. The right degrees of freedom appear by impossing the constraint
\be
F_p = (-1)^{\left[p\ov 2\right]} \star F_{10-p}\ ,
\label{reddiform}
\ee
valid in Minkowskian signature spacetimes.

\subsection{Type-IIB to massive IIA}

The Hodge duals of the RR fluxes (\ref{IIBfluxansatz}) are
\ba
&& F_7 = - (\star F_3) = -m e^{-3 A} {\rm Vol}(M_7) - e^{3 A} \star_7 G_3 \wedge {\rm Vol}(S^3)
\nonumber\\
&& \phantom{xxxxxxxxxx} = -m e^{-3 A} e^0 \wedge e^1\wedge \cdots \wedge e^6
- \star_7 G_3 \wedge e^7\wedge e^8 \wedge e^9\ ,
\\
&& F_9 = \star F_1 =e^{3 A} \star_7 G_1 \wedge {\rm Vol}(S^3)= \star_7 G_1 \wedge e^7\wedge e^8 \wedge e^9\ .
\nonumber
\ea
Also
\be
\star H = e^{3 A} \star_7 H \wedge {\rm Vol}(S^3)\ .
\ee
To present the T-dual RR fluxes we define the forms
\ba
&& L_1 = {\bf x} \cdot {\bf \hat e} = e^{-A} r dr\ ,
\nonumber\\
&& L_2 = x_7\, \hat e^8 \wedge \hat e^9 + ({\rm cyclic\ in}\ {7,8,9})
= e^{2A}\tilde B = {e^{2 A} r^3\ov r^2 + e^{4 A}} {\rm Vol} (S^2)\ ,
\\
&& L_3= \hat e^7 \wedge \hat e^8 \wedge \hat e^9 = {r^2 e^A\ov r^2 + e^{4 A}} dr\wedge {\rm Vol}(S^2)\ ,
\nonumber
\ea
where in the last step we used \eqn{fraamwe} and spherical coordinates. They obey the identities
\be
L_1\wedge L_2 = r^2 L_3 \ ,\qq L_1 = \star_3 L_2 \ , \qq L_2 = \star_3 L_1 \ .
\ee
Using the transformation
\begin{equation}
\label{omegabis}
\hat{P}=P\Omega^{-1}
\end{equation}
with $\Omega$ as in (\ref{omega})  we arrive after some algebra at
\ba
&& \hat F_0 =m  \ ,
\nonumber\\
&& \hat F_2 = - G_2 +m \tilde B  - e^A G_1\wedge L_1 \ ,
\nonumber\\
&& \hat F_4 = e^{3A} \star_7 G_3 -  e^A G_3 \wedge L_1  - G_2 \wedge \tilde B + e^{3A} G_1\wedge L_3\ ,
\nonumber\\
&& \hat F_6 = e^{3A} G_3 \wedge L_3 + e^{3A} \star_7 G_3 \wedge \tilde B
+ e^{-2 A} \star_7 G_2 \wedge L_1 - e^{3 A} \star_7 G_1\ ,
\label{flllhatIIB}
\\
&& \hat F_8 =m e^{-2 A} {\rm Vol}(M_7) \wedge L_1 - \star_7 G_2\wedge L_3 - e^A \star_7 G_1 \wedge L_2\ ,
\nonumber\\
&& \hat F_{10} = -m {\rm Vol}(M_7)\wedge L_3 = -m {\rm Vol}(M_{10})\ .
\nonumber
\ea

\no
One may check that the Bianchi identities and the flux equations of massive IIA are indeed obeyed using the
corresponding formulae for type-IIB. In doing so the identity
\be
d(e^{3A} L_3) + e^A \tilde H\wedge L_1 =0 \ ,
\ee
where $\tilde H = d\tilde B$, proves useful.
One may also check that \eqn{reddiform} is obeyed and therefore
we may use $F_p$, with $p=0,2,4$ as the independent
flux forms.

\subsection{Massive IIA to IIB}

In this case the Hodge duals of the RR fluxes (\ref{formsIIA}) are
\ba
&& F_6 = -(\star F_4) =  - e^{3 A} \star_7 G_4 \wedge {\rm Vol}(S^3) - e^{-3 A} \star_7 G_1 \ ,
\nonumber
\\
&& F_8 =\star F_2 =  e^{3 A} \star_7 G_2\wedge  {\rm Vol}(S^3) \ ,
\\
&& F_{10}=  -(\star F_0) = -m e^{3 A} {\rm Vol}(M_7)\wedge {\rm Vol}(S^3)\  .
\nonumber
\ea
Using (\ref{omegabis}) we arrive after some algebra at
\ba
&& \hat F_1 = -G_1 - m e^A L_1\ ,
\nonumber\\
&& \hat F_3 =  e^{3 A} \star_7 G_4 - e^A G_2\wedge L_1 -  G_1\wedge \tilde B + m e^{3 A} L_3\ ,
\nonumber\\
&& \hat F_5 = e^{3A} \star_7 G_4\wedge \tilde B
+ e^{3 A} G_2\wedge L_3 - e^{3 A} \star_7 G_2 - e^A G_4 \wedge L_1\ ,
\\
&& \hat F_7 = e^{3 A} G_4 \wedge L_3 + e^{-2 A} \star_7 G_1 \wedge L_1
- e^{3A} \star_7 G_2 \wedge \tilde B
+ m e^{3 A} {\rm Vol}(M_7)\ ,
\nonumber\\
&& \hat F_9 = m e^A {\rm Vol}(M_7) \wedge L_2 - \star_7 G_1\wedge L_3\ .
\nonumber
\ea
One may also check that the Bianchi identities and the flux
equations of type-IIB are indeed obeyed using the
corresponding formulae for massive IIA.
Furthermore, \eqn{reddiform} is obeyed and therefore
we may use $F_p$, with $p=1,3,5$ as the independent
flux forms.

%%%%%%%%%%%%%

\providecommand{\href}[2]{#2}\begingroup\raggedright
\endgroup

\end{document}